\documentclass[preprint2]{aastex6}

\usepackage{natbib}
\usepackage{graphicx}
\usepackage{bm}
\usepackage{amsmath}
\usepackage{amsfonts}
\usepackage{amssymb}
\usepackage{xcolor}
\usepackage{mathrsfs}
\usepackage[caption=false]{subfig}
\bibpunct{(}{)}{;}{a}{,}{,} %to follow the A&A bib style
\usepackage{ulem}
\usepackage{CJK}
\usepackage{txfonts}

\usepackage[colorlinks, linkcolor=violet, anchorcolor=green, citecolor=blue]{}

\newcommand{\HII}{H{\scriptsize II}}

\def\arcsec{$^{\prime\prime}$}

\def\purple#1 {{\textcolor{purple}{#1}}\ }
\def\red#1 {\textcolor{red}{#1}}
\def\new#1 {{\bf #1 }}

\def\blue#1 {{\textcolor{blue}{#1}}\ }
\def\sioff  {SiO $J=5\rightarrow4$}%

\def\kms {km\,s$^{-1}$}

\begin{document}
%\begin{CJK*}{UTF8}{gbsn}

\title{A SiO $J=5\rightarrow4$ Survey Toward Massive Star Formation Regions}

\author{Shanghuo Li \altaffilmark{1,2,3}} 
\author{Junzhi Wang \altaffilmark{1,4}}
\author{Min Fang\altaffilmark{5}}
\author{Qizhou Zhang \altaffilmark{2}}
\author{Fei Li\altaffilmark{1,3}}
\author{Zhi-Yu Zhang\altaffilmark{6,7}}
\author{Juan Li\altaffilmark{1}}
\author{Qingfeng Zhu\altaffilmark{8}}
\author{Shaoshan Zeng\altaffilmark{9}}

\altaffiltext{1}{Shanghai Astronomical Observatory, Chinese Academy of Sciences, 80 Nandan Road, Shanghai 200030, China}
\altaffiltext{2}{Center for Astrophysics $|$ Harvard \& Smithsonian, 60 Garden Street, Cambridge, MA 02138, USA}
\altaffiltext{3}{University of Chinese Academy of Sciences, 19A Yuquanlu, Beijing 100049, China}
\altaffiltext{4}{Key Laboratory of Radio Astronomy, Chinese Academy of Sciences, 210008, Nanjing, China}
\altaffiltext{5}{Department of  Astronomy and Steward Observatory, University of Arizona, 933 N Cherry Ave., Tucson, AZ 85721, USA}
\altaffiltext{6}{Institute for Astronomy, University of Edinburgh, Blackford Hill, Edinburgh, EH9 3HJ, UK}
\altaffiltext{7}{European Southern Observatory, Karl-Schwarzschild-Strasse 2, D-85748 Garching, Germany}
\altaffiltext{8}{Astronomy Department, University of Science and Technology, Chinese Academy of Sciences, Hefei 210008, China}
\altaffiltext{9}{School of Physics and Astronomy, Queen Mary University of London, Mile End Road, E1 4NS London, UK}

\email{shanghuo.li@gmail.com; jzwang@shao.ac.cn}
\shorttitle{}
\shortauthors{Li et al}

\begin{abstract}
We performed a survey in the \sioff\  
line toward a sample of 
199 Galactic massive star-forming 
regions at different  evolutionary stages with the 
SMT 10 m and CSO 10.4 m telescopes. The sample consists 
of 44 infrared dark clouds (IRDCs), 86 protostellar 
candidates, and 69 young \HII\ regions. We detected
\sioff\ line emission in 102 sources, with a 
detection rate of 57\%, 37\%, and 65\% for IRDCs, 
protostellar candidates, and young \HII\ regions, 
respectively. We find both broad line with Full Widths 
at Zero Power (FWZP) $>$ 20 \kms  and narrow line 
emissons of SiO in objects at various evolutionary stages, 
likely associated with high-velocity shocks 
and low-velocity shocks, respectively. 
The SiO luminosities do not show apparent differences 
among various evolutionary stages in our sample. 
We find no correlation between the SiO abundance and 
the luminosity-to-mass ratio, indicating that the
SiO abundance does not vary significantly in regions at  
different evolutionary stages of star formation. 
\end{abstract}


\maketitle

\label{firstpage}

%%%%%%%%%%%%%%%%%%%%%%%%%%%%%%
%INTRODUCTION
%%%%%%%%%%%%%%%%%%%%%%%%%%%%%%
\section{Introduction}
\label{intro}
The formation of massive stars has been studied both 
observationally and theoretically for several decades 
\citep{1979ARA&A..17..345H,2004RvMP...76..125M,
2007ARA&A..45..565M,2007ARA&A..45..339B,2018ARA&A..56...41M}.  
However, from an observational point of view, it is challenging 
to study the formation mechanism due to the large 
distance, the high extinction, and the short timescales of 
massive star formation 
\citep{2007ARA&A..45..481Z,2018ARA&A..56...41M}.
This has limited our understanding of the processes that 
regulate massive star formation. 
Massive star formation can produce strong feedback to 
their parent molecular clouds and surrounding 
interstellar medium (ISM) in form of jets, outflows, 
ultra-violet radiation, and H{\scriptsize II} regions. 
\citep{2002ARA&A..40...27C,2007prpl.conf..245A,
2015ApJ...804..141Z,2016ARA&A..54..491B,2017MNRAS.466..248L}. 
These feedbacks also provide crucial clues to 
the massive star formation processes.

Interactions between jets/outflows and surrounding medium 
have been found in massive star forming regions at 
multiple evolutionary stages --  from infrared dark 
clouds  (IRDCs) to \HII\ regions \citep{2002ARA&A..40...27C,
2007prpl.conf..245A,2007ApJ...654..361Q,2011A&A...530A.118P,
2015ApJ...804..141Z,2016ARA&A..54..491B,2016A&A...595A.122L,
2017MNRAS.466..248L}.
A number of gas tracers has been employed to study shocks 
in the interactions. 
Superior to the CO and 
HCO$^{+}$ lines which are often contaminated by the 
emission from the ambient material, 
the rotational transitions of Silicon monoxide (SiO) 
have been widely used in studying shock conditions, 
especially those associated with protostellar jets and outflows 
\citep{1999A&A...345..949C,
1999ApJ...527L.117Z,2013A&A...550A.106L,
2014A&A...570A..49L,2013A&A...550A..81C,
2018A&A...620A..31M}.
This is because that shock activities release Si atoms from
dust grains into gas phase via sputtering or vaporisation
\citep{2008A&A...482..809G,2008A&A...490..695G}. 
Then SiO molecules are formed dominantly by 
the reaction between Si and O$_{2}$ and OH in the 
gas phase 
\citep{1989A&A...222..205H,1997A&A...322..296C,
1997A&A...321..293S,2001A&A...372.1064L,
2014CPL...610..335R}. Alternatively, SiO can 
also be formed via less prominent routes that 
involve Si$^{+}$ (OH, H), SiO$^{+}$ (H$_{2}$, H) 
and HSiO$^{+}$ (e, H) \citep{2013A&A...554A..35L}.

The SiO abundance in molecular outflows can be 
enhanced  by up to six orders of magnitude relative 
to that in the quiescent regions 
\citep{1992A&A...254..315M,2007MNRAS.380..246G,
2011A&A...526L...2L}. 
This makes SiO an excellent tracer for investigating shock 
activities in star formation regions. 
SiO has been used to trace shocks with different 
velocities (low-velocity: $\varv \leqslant$ 10 \kms, 
medium-velocity: $\varv \sim 10 - 20$ \kms, high-velocity:
$\varv \sim 20 - 50 $\kms , and extremely high-velocity: 
$\varv \sim$ 100 \kms).
The low- and medium-velocity shocks could be associated 
with outflows \citep{2017MNRAS.466..248L}, converging 
flows \citep{2010MNRAS.406..187J,2013MNRAS.428.3425H} 
and cloud collision  \citep{2016A&A...595A.122L}. 
The high-velocity shocks, on the other hand, are most likely 
associated with jets and molecular outflows driven by embedded 
young stars \citep{2007ApJ...654..361Q}. The extremely 
high-velocity shocks have been found in the Central Molecular 
Zone (CMZ) of our  Galaxy, which is likely associated with 
larger-scale shocks in the CMZ 
\citep{1997ApJ...482L..45M,2012MNRAS.419.2961J}.

Observations of the SiO emission toward various evolutionary 
stages of star formation would allow us to characterize 
the variation in SiO with  evolution.
Observations of the SiO $J=2\rightarrow1$ line toward a 
sample of massive star formation regions by 
\citet{2010ApJ...714.1658S} found that mid-infrared bright 
sources tend to have a lower SiO abundance than that 
of mid-infrared dark sources, indicating that 
the SiO abundance decreases in more evolved massive 
star formation regions. 
A similar decreasing trend in the SiO abundance with 
time has been  reported by several studies of samples at 
different evolutionary stages 
\citep{2006A&A...460..721M,2007A&A...476.1243M,
2013A&A...557A..94S,2011A&A...526L...2L}. 
On ther other hand, other studies found a different trend 
that the SiO abundance in infrared-bright 
clumps is  higher than the one in infrared-quiet clumps 
\citep[e.g.,][]{2014A&A...563A..97G,2014A&A...562A...3M}, 
or the SiO abundance does not vary significantly among 
different evolutionary stages of massive star formation 
\citep{2012ApJ...756...60S,2014A&A...570A..49L,
2016A&A...586A.149C}. Therefore, it is still under 
debate if the SiO abundance varies with the 
evolutionary stages of massive star formation.

The SiO 5-4 transition is an excellent tracer to study 
strong shock activities that are related to 
outflows/jets driven by embedded massive protostars, 
since it requires higher excitation conditions 
(a critical density n$_{cr} \sim 1.7 \times 10^{6}\, 
\rm cm^{-3}$, upper energy level E$_{u}$ = 31.26 K) 
than those of CO, HCO$^{+}$ and SiO low $J$ transitions. 
In this work, we report a survey in the SiO~5-4 line 
of 199 massive clumps at different stages of massive 
star formations, e.g., IRDCs, protostellar candidates, 
and \HII\ regions, with the SMT 10 meter and CSO 10.4 
meter telescopes. First we  describe the sample selection 
and observations in $\S$ 2. Then, we present the data 
analysis and results in $\S$ 3.  We discuss the 
observational results in $\S$ 4, and summarize 
the main findings in $\S$ 5.

%%%%%%%%%%%%%%%%%%%%%%%%%%%%%%
%THE SAMPL  AND OBSERVATIONS
%%%%%%%%%%%%%%%%%%%%%%%%%%%%%%

\section{Samples and observations}
\label{sec:observations}
\subsection{The sample}
Our targets are selected from several catalogs in the 
literature that can be grouped into three evolutionary 
stages of massive star formation:  IRDCs, protostellar 
phase and \HII\ regions. The detailed information 
of the sample is described as follows:

\textbf{IRDCs:} 
We selected 44 massive clumps from 
\cite{2008ApJ...678.1049S}, 
which are identified as silhouettes against the bright 
Galactic background in the 8 $\mu m$ band of the 
Midcourse Space Experiment (MSX).
These objects are within a distance of 4.3 kpc \citep{2005ApJ...633..535B}. 
Among this IRDC sample, 32 sources were observed in the Spitzer\, 
24 $\mu m$ band, and 25 of them show  point-like sources in the  
24 $\mu m$ images \citep{2008ApJ...678.1049S}, indicating star 
formation in these regions. Since massive IRDCs are 
cold ($<$ 20~K) and  dense ($>$ 10$^5$~cm$^{−3}$) 
\citep{2006A&A...450..569P,2007ARA&A..45..339B}, 
they are considered the precursors of massive stars and 
star clusters.

\textbf{Protostars:} A subset (86) of  our sample  are 
massive clumps associated with 
protostellar candidates (protostars), which are originally 
selected from the \textit{IRAS} point source catalog. Among them, 
38 sources are retrieved from \cite{2005ApJ...625..864Z}, 19 
sources from \cite{2002ApJ...566..931S}, and 29 sources from 
\cite{2015MNRAS.453..645M} and \cite{2013ApJS..208...11L}. 
The  distances of these sources lie in the range of 0.7 
to 8.7 kpc \citep{2005ApJ...625..864Z,2002ApJ...566..931S,
2015MNRAS.453..645M,2013ApJS..208...11L,2014ApJ...783..130R}. 
Eighty-one sources out of 86 protostars exhibit CO outflow 
signatures  \citep{2005ApJ...625..864Z,2002ApJ...566..931S,
2015MNRAS.453..645M,2013ApJS..208...11L}.  Without associated 
\HII\ regions \citep{1974Ap&SS..27....3M,1979A&AS...35...23A,
1994ApJS...90..179G,1994ApJS...91..111W}, this subsample 
is believed to be at the intermediate evolutionary stage 
between IRDCs and  \HII\ regions \citep{2002ApJ...566..931S,
2005ApJ...625..864Z,2015MNRAS.452..637M}.

\textbf{H}\textsc{ii} \textbf{regions:} 
Sixty-nine massive clumps are selected from the 
UC \HII, C\HII\ and \HII\ region catalogs 
\citep{2003ApJS..149..375S,
2015ApJ...802...40L,2015MNRAS.453..645M,
2013ApJS..208...11L}. 
The  distances of these sources 
lie in the range of 1.2 to 7.6 kpc. 
Most of them (43) 
are associated with H$_{2}$O maser emission 
\citep{1988A&AS...76..445C,2003ApJS..149..375S}. 
The sources in this subsample are considered in 
the most advanced evolutionary stage in our sample.

\subsection{Observations}

\subsubsection{CSO 10.5 m observations}
Sixteen sources were observed in the SiO 5-4 transition at a rest 
frequency of 217.104984 GHz using the Caltech Submillimeter 
Observatory's (CSO)\footnote{http://cso.caltech.edu/} 10.4 m 
telescope in 2013 (see Table \ref{tab:obs}). In order to achieve 
good spectral baselines, 
we used the standard position switching mode with the `clean'  
reference position (OFF source) 30 arcmin  offset from 
the source in azimuth. The typical on-source time is about 
10 minutes for each target. 
The Fast Fourier Transform Spectrometer provides a 4 GHz window 
of double-side band (DSB) coverage with a velocity resolution of 
0.27 MHz (0.37 km s$^{-1}$ at 217 GHz). The SiO 5-4 transition was 
covered in the lower-side band (LSB). Typical system temperatures 
range from 250 to 400 K, and the typical rms noise 
($\delta_{T_{\rm mb}}$) is between 0.03 and 0.14 K at a velocity 
resolution of 0.37 km s$^{-1}$. 
The full-width half-maximum (FWHM) beam size is about 
27.4\arcsec\ at the observing frequency of 217.1 GHz. 
Data reduction was performed with the CLASS package, 
which is a part of the  
GILDAS\footnote{http://www.iram.fr/IRAMFR/GILDAS} software. 
The data were converted from the antenna temperature, 
$T_{A}^{*}$, to the main-beam temperature, $T_{mb}$, 
using $T_{mb}$ = $T^{*}_{A}/\eta_{\rm mb}$,  where 
$\eta_{\rm mb}$ = 0.69 is the beam efficiency measured 
by the CSO.
More detailed information of observations and data 
reduction is described in \cite{2015ApJ...802...40L}.

\subsubsection{SMT 10 m observations}
Additional observations toward 185 objects were obtained in 
the SiO 5-4 transition using the Heinrich Hertz Submillimeter 
Telescope  (SMT)\footnote{http://aro.as.arizona.edu/index.htm} 
10 m telescope in 2015 and 2016 (see Table \ref{tab:obs}). 
The standard position switching mode with the `clean'  
reference position 30 arcmin  offset from the source 
in azimuth was used for all observations. 
Pointing was checked every hour using planets 
or a known bright source close to the target when 
planets were not available. 
Typical on-source time for each source is about 30 minutes. 
The SIS ALMA Type 1.3mm receiver and Forbes Filter Bank (FFB) 
backend were employed with two filterbanks configured to 1 MHz 
(1.38 km s$^{-1}$ at 217.1 GHz) and 250 KHz (0.35 km s$^{-1}$ at 
217.1 GHz) spectral resolution sections that provide a 
bandwidth of about 1000 MHz and 50 MHz, respectively.  
The FWHM beam size is $\sim$ 28.5\arcsec\ at the observing 
frequency of 217.1 GHz. 
Data reduction was undertaken with the CLASS package. We have 
filtered spectra that show sinusoidal fluctuations, bad 
baseline and/or strong spikes in some channels. 
The data were converted from $T_{A}^{*}$ to $T_{mb}$ with 
the beam efficiency of 0.74 measured by the SMT. The 
typical system temperatures are between 150 and 250 K, 
and the typical rms noise level ($\delta_{T_{\rm mb}}$) 
is about 7 mK at the velocity resolution of 
1.38 km s$^{-1}$.  In few cases (5\%), the system temperatures 
reached as high as 350 K due to bad weather conditions.

\subsubsection{Infrared bolometric luminosity and dust 
continuum emission}
The  infrared bolometric luminosity of our sources is  
adopted from the Red MSX Source 
(RMS\footnote{http://www.ast.leeds.ac.uk/RMS/}) survey 
\citep{2013ApJS..208...11L}. The dust continuum emission 
of our sources at 870 and 850 $\mu$m is taken from the 
ATLASGAL survey 
\citep{2009A&A...504..415S} and SCUBA Legacy survey 
\citep{2008ApJS..175..277D}, respectively. These data were 
used to estimate the gas mass and H$_{2}$ column density.

%%%%%%%%%%%%%%%%%%%%%%%%%%%%%%
%RESULTS
%%%%%%%%%%%%%%%%%%%%%%%%%%%%%%
\section{Results and analysis}
\label{sec:results}
The source parameters from the literature are listed in 
Tables  \ref{tbl:sio} and \ref{tbl:no-sio}. The WISE 22 $\mu$m fluxes retrieved from 
the WISE archive are listed in column 17. The source temperature 
estimated from the NH$_{3}$ data and its reference are listed 
in column 18. The source distance, classification, reference 
and the telescope used are listed in columns 19, 20, 21 and 
22 of Table \ref{tbl:sio}, respectively. 
The sources without SiO 5-4 line detection are summarized in 
Table \ref{tbl:no-sio}.

\begin{figure*}[!tb]
	\epsscale{0.55}
	\plotone{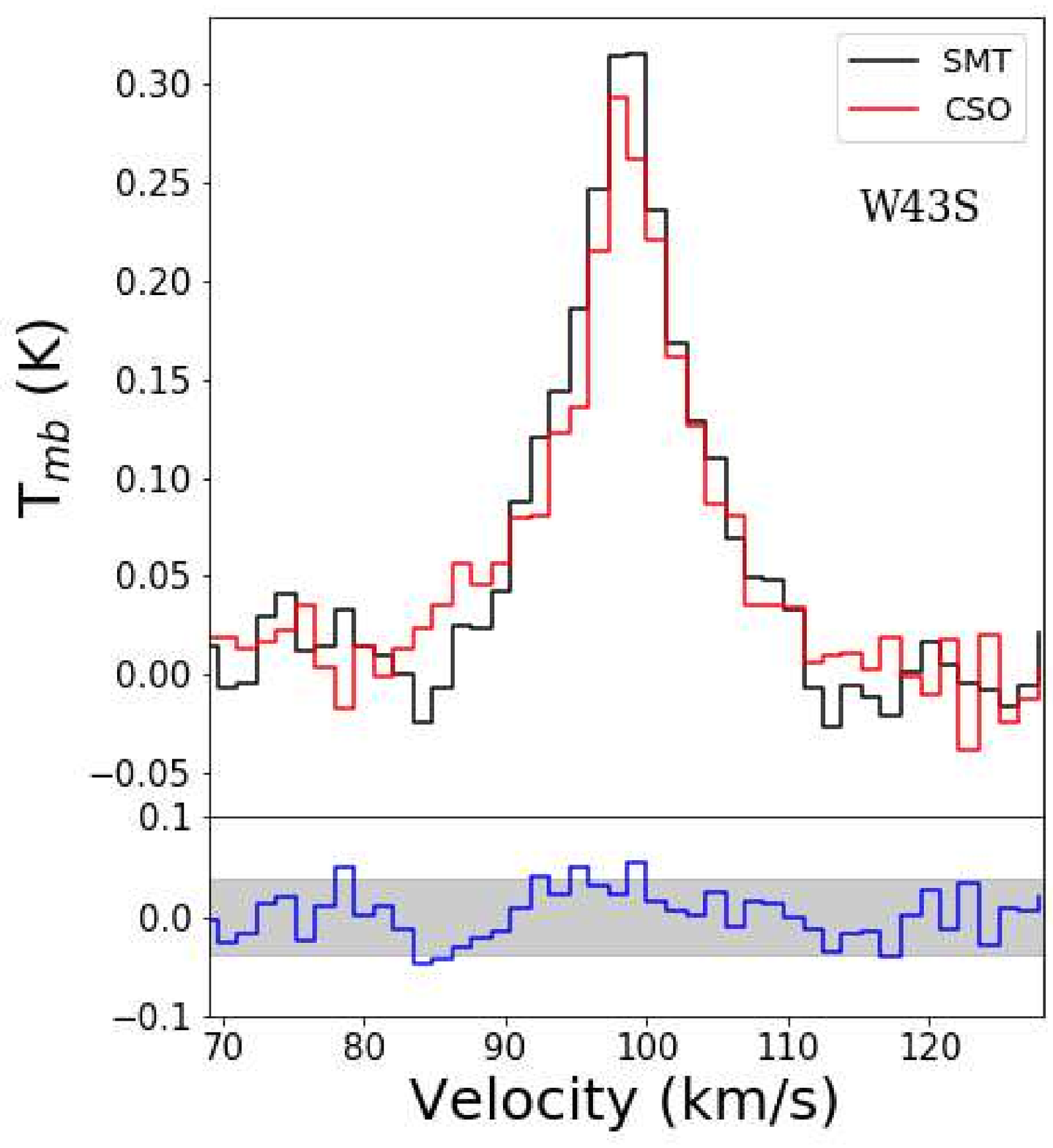}
	\plotone{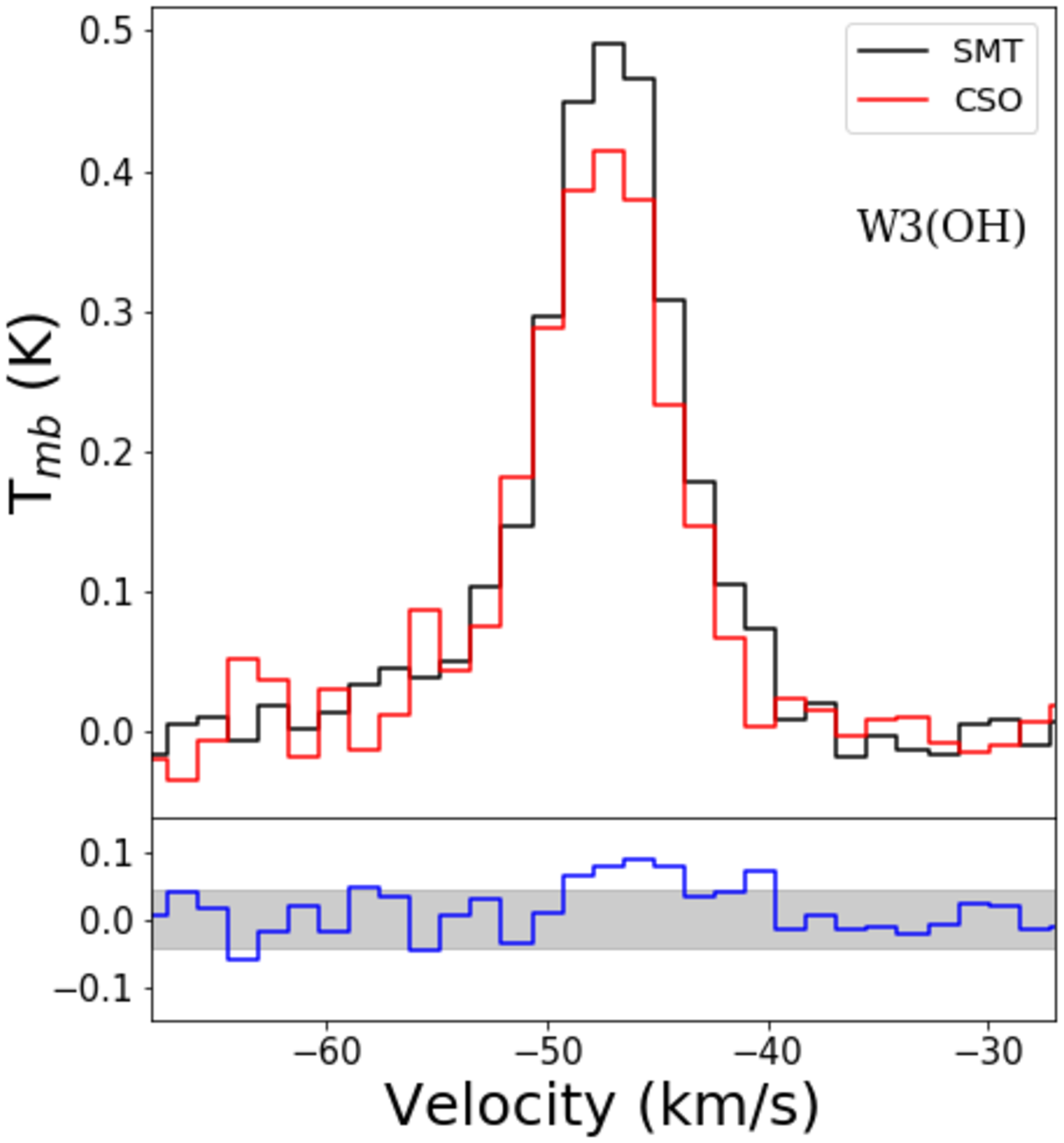}
	\caption{Spectral lines of the SiO 5-4 transition (black 
	and red histograms) obtained from the observations with 
	the SMT and CSO telescopes, respectively. 
    The bottom panels show the residuals after the SMT data 
    being subtracted by the CSO data, while the gray shadow 
    shows the rms level of the spectral line observed by 
    the CSO. The spectral line data from the CSO are 
    re-sample to 1.38 km s$^{-1}$, the same spectral 
    resolution of the SMT. 
    \textit{Left:} The spectra of the SiO 5-4 line in W43S.
    \textit{Right:} The spectra of the SiO 5-4 line in W3(OH).}
\label{fig:com}
\end{figure*}

In order to check the consistency between observations of 
SMT and CSO, we observed W43S and W3(OH) with the 
SMT and CSO telescopes. Figure \ref{fig:com} shows the 
comparison of line profiles of the SiO 5-4 transition. 
The CSO spectra have been resampled to the same spectral 
resolution of the SMT spectrum. The observed spectra 
from CSO and SMT observations show  similar line profiles. 
The difference in the velocity integrated intensity 
between two observations is 8\% and 3\% for W3(OH) 
and W43S,  respectively. These differences are within the
1$\sigma$ uncertainty. This suggests that there 
are no significant discrepancy between the SMT and 
CSO observations.

\subsection{Detection rate}
\label{det-rate}
Since the majority of the SiO 5-4 spectra show a non-Gaussian profile 
(see Figure \ref{fig:spec}), we used the Full Width at Zero 
Power (FWZP) velocity range  to calculate the integrated 
intensity (gray in Figure \ref{fig:spec}), 
$\int T_{\rm mb}\, dv$. The FWZP is visually identified 
where the emission higher than the 1$\sigma$ rms noise level. 
The rms of the velocity integrated intensity is  
$\sigma_{area}$ = $\sqrt{N} \sigma \Delta v$, 
where $\Delta v$ is the velocity resolution,  
$\sigma$ is the rms noise level of spectra, 
and $N$ is the number of channel within the integrated velocity range. 
The derived parameters are summarized in Table \ref{tbl:sio}, 
including the line central 
velocity ($\rm V_{\rm LSR}$), velocity integrated intensity 
($\int T_{\rm mb}\, dv$), line peak brightness temperature ($T_{\rm mb}$), 
and FWZP.

The SiO 5-4 emission was detected in 102 out of the 199 sources 
with velocity integrated intensity above 3$\sigma_{area}$.
The detection rate is about 51\% for the entire sample. 
Among them, the SiO 5-4 emission was detected in  25 out 
of the 44 IRDCs with a detection rate of 57\%, 32 out 
of the 86 protostars with a detection rate of 37\%, 
and 48 out of 69 \HII\ regions with a  detection 
rate of 65\%.   High detection rate in IRDCs indicates 
that shock activities are common even in the early 
evolutionary stage of star formation. 
The highest detection rate in \HII\ regions indicates 
that their shocked activities are more common than 
that in relatively less evolved sources.  
Even though the CO outflow detection rate is  94\% 
in protostars, they have low detection rate in 
SiO emission (37\%). This could partly due to the fact that the 
\sioff\ transition requires a much higher excitation condition  
than that of the CO low J transitions. 
On the other hand, we can not rule out that the low detection 
rate is due to limited sensitivity of observations since the SiO  
lines emissions are much fainter than the CO lines in these 
star forming regions.

\begin{figure*}[!tb]
	\epsscale{1.2}
	\plotone{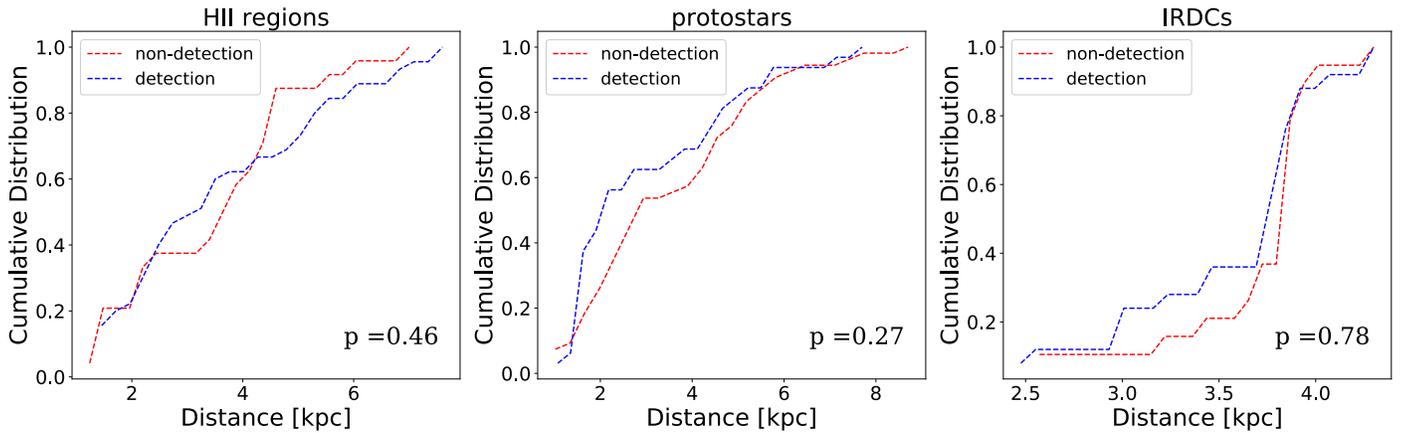}
	\caption{Distance distributions for sources with and 
	without SiO detections (The blue and red dash lines). 
	From \textit{left} to \textit{right} are \HII\ regions, 
	protostars and IRDCs. 
    The $p$-value from the KS test is presented inside the panel. 
    }
	\label{fig:comp1}
\end{figure*}

The large range of source distances from 0.7 to 8.7 kpc 
in this sample may introduce a bias in the detection 
rate. In order to examine the effect of distance, 
we computed the detection rate for 155 sources 
(44 IRDCs, 62 protostars and 48 \HII) as a distance 
limited sample (D $\leqslant$ 4.4 kpc). 
The detection rates are 57\%, 40\% and 63\% 
for IRDCs, protostars, and \HII\ regions, respectively. 
This is consistent with the results of the whole sample.  
In addition, we employed the Kolmogorov-Smirnov (KS) test 
to compare the distance distributions between sources that 
show detection and non-detection of SiO. 
The p-value of KS test is a probability that measures the 
evidence against the null hypothesis. 
If the p-value is much smaller than 5\%, we can reject the null 
hypothesis that two samples are draw from same parent 
distribution, while the null hypothesis can not be ruled out
if the p-value much larger than 5\%. 
The KS test reveals 
that the p-value is 46\%, 27\% and 78\% for HII regions, 
protostars and IRDCs, respectively. 
This indicates that the distance difference between 
detection and non-detection sources is no statistically 
significant for all three categories. 
Figure \ref{fig:comp1} shows the cumulative distribution 
of source distance.

\subsection{Line width}
\label{line_profile}
Based on the derived FWZP, we divided the SiO spectra into 
four groups, (1) low-velocity,  FWZP $\leqslant$ 10 
km s$^{-1}$; (2) intermediate-velocity, 
10 ${\rm km\, s^{-1}} <$ FWZP $\leqslant$ 20 \rm km\, 
s$^{-1}$; (3) high-velocity, 20 ${\rm km\, s^{-1}} <$ 
FWZP $\leqslant $ 50 km s$^{-1}$; (4) very high-velocity, 
FWZP $>$ 50 $\rm km\, s^{-1}$, similar to the criteria in 
\cite{2007ApJ...668..348B}.  
There are nine sources in the low-velocity regime, 43 sources in 
the intermediate-velocity regime, 47 sources in the high-velocity 
regime and three sources in the very high-velocity regime.
Note that the majority of sources (91\%) are associated with 
FWZP $>$ 10 km s$^{-1}$, which indicates that these SiO 
emissions arise from gases of relatively high-velocities 
with respect to the ambient gas velocity. 
Table \ref{tab:vel} lists the number of sources in the 
four groups for the three categories.

As shown in Figure \ref{fig:spec}, most of the sources exhibit 
an asymmetric line profile that can not be fitted by a 
single Gaussian profile, while some of the sources show 
extended line wings and 
12 sources (05358+3543, G18151-1208MM1, G18151-1208MM2, 
I18223-1243MM4, G023.60+00.00MM3, G024.33+00.11MM5, 
S87, ON2S, G97.53+3.19, 23385+6053, G5.89-0.39 and DR21S) 
exhibit two distinguishable velocity components. 
The non-Gaussian 
extended line wings in the SiO 5-4 emission were detected 
in 6 out of 25 IRDCs, 12 out of 32 protostars and 14 out of 
48 \HII\ regions.

\begin{figure*}[!tb]
	\epsscale{1.3}
	\plotone{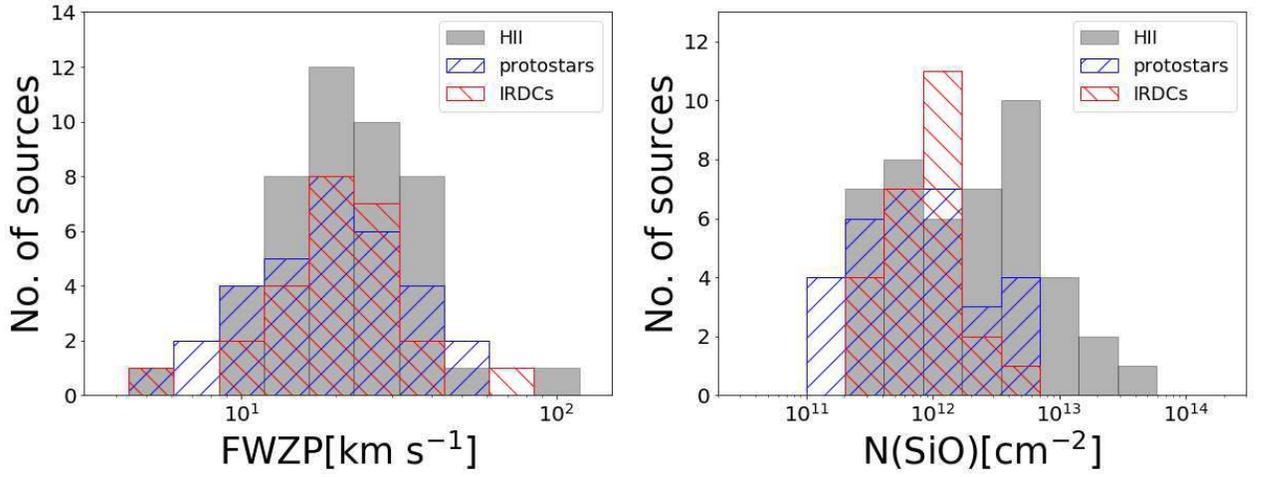}	
	\caption{\textit{Left:} Histograms showing the FWZP 
	distributions for \HII\ regions (gray), protostars 
	(blue), and IRDCs (red). 
    \textit{Right:} Same as left but for SiO column 
    density distributions. 
    }
	\label{fig:width}
\end{figure*}

The distributions of FWZP for IRDCs, protostars and \HII\ 
regions are presented in Figure \ref{fig:width}. From these 
distributions, we note that there are no significant differences 
between these three categories. The mean FWZP is 20 km s$^{-1}$, 
19 km $s^{-1}$, and 20 km s$^{-1}$ for IRDCs, protostars, and 
\HII\ regions, respectively. Furthermore, we 
derived the median FWZP for all three categories, which is  
23 km s$^{-1}$, 22 km $s^{-1}$, and 24 km s$^{-1}$ for IRDCs, 
protostars, and \HII\ regions, respectively. The statistical 
mean FWZP as well as the median FWZP is similar among 
different evolutionary stages. 
Further discussions of relationship between FWZP 
and evolutionary stages are presented in Section 
\ref{SiO_evol_fwzp}.

\subsection{SiO luminosity}
\label{luminosity}
Using the integrated intensity $\int T_{\rm mb}\, dv$ and 
the source distance, we estimated the SiO 5-4 luminosity
($L_{\rm SiO}$) within the beam (hereafter, 
luminosity) through \citep[see][]{2013ApJ...775...88N}: 
\begin{eqnarray}
\label{Lum_SiO}
L_{\rm SiO} &=& 4\pi \times d^{2} \times \int T_{\rm mb}\, dv \nonumber \\
			& &\simeq  {\rm 2.3 \times 10^{-4}\ L_{\odot}}  \times \left( \frac{d}{\rm 6\,kpc} \right)^2   
			\frac{\int T_{\rm mb}\, dv}{\rm1\, K\, km\, s^{-1}}
\end{eqnarray}

The derived $L_{\rm SiO}$ is between 1.3$\times 10^{-5}$ and 
1.4$\times 10^{-4}\ L_{\odot}$ for IRDCs, between 
8.5$\times 10^{-7}$ and 2.5$\times 10^{-4}\ L_{\odot}$ for 
protostars, and between 3.6$\times 10^{-6}$ and 
3.7$\times 10^{-3}\ L_{\odot}$ for \HII\ regions. 
The mean value is 4.9$\times 10^{-5}$, 
4.8$\times 10^{-5}$ and 3.6$\times 10^{-3}$ $L_{\odot}$ for IRDCs, 
protostars and \HII\ regions, respectively. The median value 
is 3.8$\times 10^{-5}$, 2.3$\times 10^{-5}$ and 
1.3$\times 10^{-4}$ $L_{\odot}$ for IRDCs, protostars and \HII\ 
regions, respectively.  The derived $L_{\rm SiO}$ is summarized in 
Table \ref{tbl:sio}.

\begin{figure*}[!tb]
	\epsscale{1.3}
	\plotone{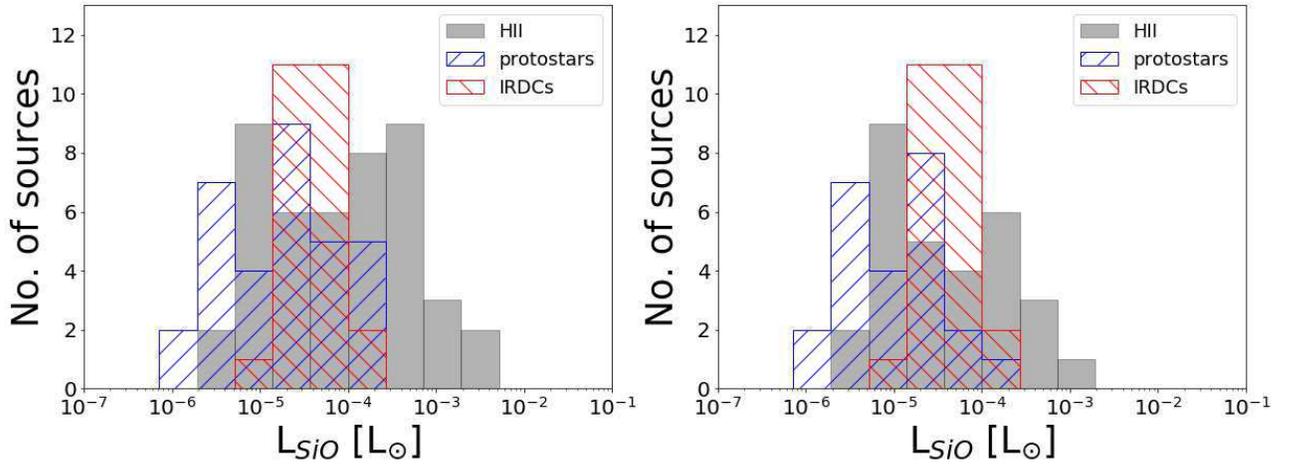}
	\caption{\textit{Left:} Histograms showing the 
	$L_{\rm SiO}$ distributions of IRDCs (red), protostars 
	(blue) and \HII\ regions (black).
    \textit{Right:}  Same as left but for distance limited 
    samples.}
	\label{fig:lum}
\end{figure*}

Figure \ref{fig:lum} shows distributions of $L_{\rm SiO}$ for 
the three categories. It is clear that 
there is a fraction of \HII\ regions (40\%) associated with 
relatively higher $L_{\rm SiO}$ ($>$ 2.5$\times 10^{-4}\ L_{\odot}$)
than that of IRDCs and protostars. 
The $L_{\rm SiO}$ of IRDCs is concentrated around 6 
$\times 10^{-5}\ L_{\odot}$, while the protostars are distributed 
throughout between 10$^{-6}$ and 3$\times 10^{-4}\ L_{\odot}$. 
We also plotted the $L_{\rm SiO}$ distributions for a distanced 
limited sample in the right panel of Figure \ref{fig:lum}, 
which shows distributions similar to those of the entire sample. 
There is still a fraction of \HII\ regions (30\%) associated 
with relatively higher $L_{\rm SiO}$ 
($>$ 2.5 $\times 10^{-4}\ L_{\odot}$) in the distance limited 
sample, which indicates that the high $L_{\rm SiO}$ in \HII\ 
regions is likely not due to the large distances (see Section 
\ref{dis:lu_Nsio} for a discussion on $L_{\rm SiO}$).

\subsection{SiO column density}
\label{column}
Assuming a  beam-filling factor of unity and a local 
thermodynamic equilibrium (LTE) condition for the
SiO 5-4 emission, the beam-averaged column density 
(hereafter, column density)  can be calculated from 
the velocity integrated intensity 
$\int T_{\rm mb}\ dv$ with the following equation: 
\begin{eqnarray}
\label{N_SiO1}
N &=& \frac{3h}{8\pi^{3} S \mu^{2}} 
\frac{Q_{\rm rot}(T_{ex}) }{ g_{J}\, g_{K}\, g_{s}} 
\frac{exp\left(\frac{E_{\rm u}}{k T_{\rm ex}}\right)}
{exp\left(\frac{h \nu}{k T_{\rm ex}}\right) - 1}
\nonumber \\ 
&&\times 
\frac{1}{J_{\nu}(T_{\rm ex}) - J_{\nu}(T_{\rm bg})} 
\frac{1}{1 - exp(-\tau)} \int T_{\rm mb}\, dv 
\end{eqnarray}
where $J_{\nu}(T)$ = $\frac{ h\nu/k}{exp( h\nu/kT) - 1}$, 
$h$ is the Planck constant, 
$k$ is the Boltzmann constant, 
and $T_{\rm ex}$  and $T_{\rm bg}$ are the excitation temperature 
of the gas and the temperature of the background radiation, 
respectively. 
The line strength $S$ = $J_{u}/(2J_{u} + 1)$, the dipole moment 
$\mu$ = 3.1 D for the SiO, line frequency $\nu$ = 217.104 GHz, 
and the upper level energy E$_{\rm u}/k$ =  31.26 K are taken 
from the splatalogue 
catalog\footnote{http://www.cv.nrao.edu/php/splat/}. 
The rotational degeneracy $g_{J}$ is equal to 2$J_{u}$ + 1,
while the K degeneracy g$_{\rm K}$  and nuclear spin 
degeneracy g$_{\rm s}$ are equal to 1. The partition 
function Q$_{\rm rot}(T_{ex})$ is approximated as 
$kT_{\rm ex}$/$h$B + 1/3, where 
B = 21711.967 MHz is the rotational constant taken from the
JPL catalog\footnote{http://spec.jpl.nasa.gov/ftp/pub/catalog/catdir.html} 
\citep{1998JQSRT..60..883P}. 
Adopting  $T_{\rm bg}$ of 2.73 K and assuming that the SiO 5-4 
line is optically thin ($\tau \ll$ 1), equation (\ref{N_SiO1}) 
can be rewritten as:
\begin{eqnarray}
\label{N_SiO2}
N(\rm SiO) 
&&= 1.6 \times 10^{11}\, {\rm cm^{-2}} \times 
\frac{(T_{\rm ex}+0.35) \, exp\left(\frac{31.26}{T_{\rm ex}} \right)}
{exp(\frac{10.4}{T_{ex}}) - 1} 
\nonumber\\
&& \frac{1}{J_{\nu}(T_{\rm ex}) - J_{\nu}(T_{\rm bg})} 
\int T_{\rm mb}\, dv
\end{eqnarray}
In order to understand the effect of excitation 
temperatures on the SiO total column density, a wide 
range of excitation temperatures, from 10 K to 100 K, 
has been used to calculate the column density. 
We found that the estimated SiO column densities agree 
within a factor of two in this temperature range. 
This indicates that the SiO 5-4 column density does not 
strongly depend on the temperature, which is consistent 
with that reported in \cite{2007A&A...462..163N}. 
In this paper, we use the 
kinematic temperatures estimated from the NH$_{3}$ data 
to calculate the SiO column density 
\citep{2011MNRAS.418.1689U,2012A&A...544A.146W}. 
%\new{
For sources where the NH$_{3}$ measurements are not 
available,  
%}
we adopt an excitation temperature $T_{\rm ex}$ of 18 K, 
25 K and 30 K for IRDCs, protostars and \HII\ regions, 
respectively.

The derived SiO column densities range from 
2.8$\times 10^{11}$ to 3.9$\times 10^{12}$ cm$^{-2}$ for 
IRDCs, from 9.6$\times 10^{10}$ to 4.7$\times 10^{12}$ cm$^{-2}$ 
for protostars, and from 2.4$\times 10^{11}$ to  2.9$\times 10^{13}$ 
cm$^{-2}$ for \HII\ regions (Table \ref{tab:NSiO}). 
The median and mean column densities are  1.0$\times 10^{12}$ 
and 9.0$\times 10^{11}$ cm$^{-2}$ for IRDCs, 
1.2$\times 10^{12}$ and 6.1$\times 10^{11}$  for protostars, 
and 4.4$\times 10^{12}$ and 2.2$\times 10^{12}$ cm$^{-2}$ for 
\HII\ regions. The results are comparable to the measurement of 
SiO column densities toward other massive star formation regions 
\citep[e.g.,][]{2010ApJ...714.1658S,2014A&A...562A...3M,
2016A&A...586A.149C}. 
Figure \ref{fig:width} presents the distributions of the SiO 
column densities for three categories. There are no significant  
differences between IRDCs and protostars, while the values 
in \HII\ regions are relatively higher  than that in both 
IRDCs and protostars. 
The computed SiO column densities are 
listed in Column 9 of Table \ref{tbl:sio}.

\subsection{Gas mass and H$_{2}$ column density}
\label{Mgas}
We used the emission of dust at 870 or 850 $\mu$m 
retrieved from the ATLASGAL and SCUBA Legacy survey 
\citep{2009A&A...504..415S,2008ApJS..175..277D} to 
estimate the gas mass with the formula:
\begin{equation}
\label{dust_mass}
M_{\rm gas}=\eta \frac{S_{\nu}\, d^{2}}{B_{\nu}(\rm T)\kappa_{\nu}}
\end{equation}
where $M_{\rm gas}$ is the gas mass, $\eta$ = 100 is 
the gas-to-dust ratio, d is the source distance, 
$S_{\nu}$ is the integrated continuum  flux at a frequency 
of $\nu$, $B_{\nu}(\rm T)$ is the Planck function 
at a temperature of $\rm T$, and $\kappa_{\nu}$ 
is the dust opacity at a frequency of $\nu$. We adopt 
$\kappa_{\nu}$ = 10(${\nu}$/1.2 THz)$^{\beta}$ cm$^{2}$  
g$^{-1}$ \citep{1983QJRAS..24..267H} and assumed 
$\beta$ = 1.5. 
The derived masses range from 45 up to 17038 M$_{\odot}$, 
which is summarized in Column 15 of Table \ref{tbl:sio}.

We also estimate the peak H$_{2}$ column density using 
the dust emission: 
\begin{equation}
\label{H2_col}
N(\rm H_{\rm 2})=\eta \frac{F_{\nu}}{B_{\nu}(\rm T) \; \Omega_{\rm A} \; \mu_{\rm H_{\rm 2}} \; \rm m_{\rm H} \; \kappa_{\nu}}
\end{equation}
where $\rm F_{\nu}$ is the peak flux density, 
$\mu_{\rm H_{\rm 2}}$ ($\sim$ 2.8) is the molecular weight per 
hydrogen molecule \citep{2008A&A...487..993K}, $\rm m_{\rm H}$ 
is the atomic mass unit, 
$\Omega_{\rm A}$ is the beam solid angle, which is 
$\Omega_{\rm A}$ = ($\pi \theta^{2}_{\rm HPBW}$/4ln(2)) 
for a Gaussian beam.
In order to be consistent with the SiO 5-4 data, the dust 
continuum  have been convolved with a beam size of 29$^{''}$.  
The derived $\rm H_{\rm 2}$ column densities are between 
6.4 $\times$ 10$^{21}$ and 6.0 $\times$ 10$^{23}$ cm$^{-2}$, 
which are listed in Column 16 of Table \ref{tbl:sio}.

\subsection{Bolometric luminosity to mass ratio} 
The bolometric luminosity-to-mass ratio ($L/M$) is thought 
to be a good indicator of the evolutionary stage of star 
formation. 
A low $L/M$ value corresponds to an early 
evolutionary stage of star formation, and vice versa 
\citep{2008A&A...481..345M}. 
We have retrieved the bolometric luminosity ($L_{\rm bol}$) 
from the RMS survey  for the sources that show SiO emission 
\citep{2013ApJS..208...11L}. 
Forty-one \HII\ regions, 26 protostars and 6 IRDCs have 
bolometric luminosity information (Table \ref{tbl:sio}).

\begin{figure*}[!tb]
	\epsscale{1.2}
	\plotone{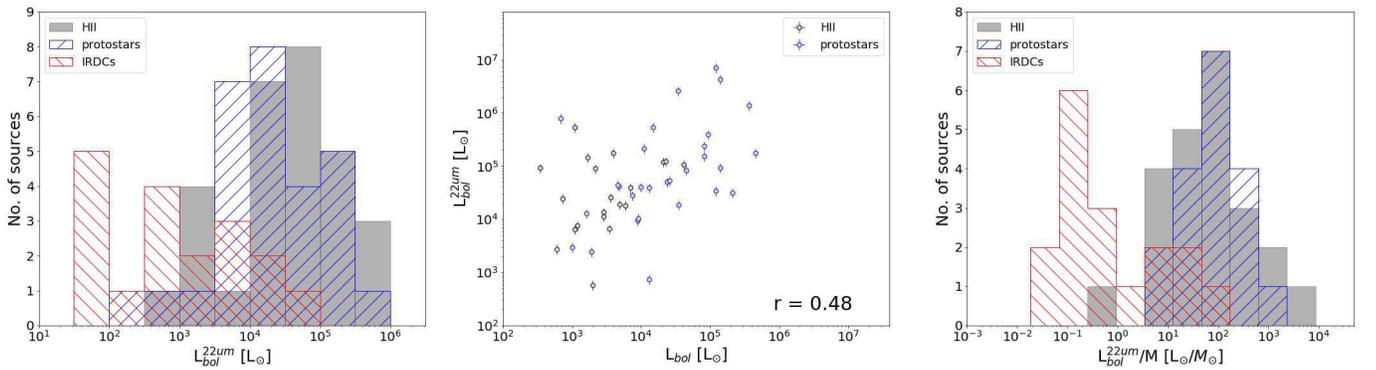}
	\caption{\textit{Left:} Histograms are the distributions 
	of bolometric luminosity derived from WISE 22 $\mu$m for 
	three categories. 
	\textit{Middle:} The bolometric luminosity derived from WISE 
	WISE 22 $\mu$m versus bolometric luminosity derived by SED fitting. 
	Blue and gray circles represent protostellars and \HII\ regions, 
	respectively.  
	\textit{Right:} Histograms are the distributions of 
	$L_{\rm bol}^{\rm 22um}/M$  for three categories. 
    }
	\label{fig:L22um}
\end{figure*}

However, the number of IRDCs that have bolometric luminosity 
measurements is not sufficient for a statistical study. 
Thus, we included the sources (20 IRDCs, 28 protostars, 
and 33 HII regions) that have 22 $\mu$m emission for a 
statistical study of the properties of SiO 
in different evolutionary stages 
\citep[see, e.g.,][]{2016A&A...586A.149C}. 
For consistency, we use the WISE 22 $\mu$m emission to 
estimate the bolometric luminosity ($L_{\rm bol}^{22 \mu m}$) 
with the scaling factor derived by \cite{2011A&A...525A.149M}. 
The distribution of  $L_{\rm bol}^{22 \mu m}$ is shown in 
Figure \ref{fig:L22um}. Note that the $L_{\rm bol}^{22 \mu m}$ 
tends to be higher in more evolved sources.

Figure \ref{fig:L22um} shows a comparison of 
$L_{\rm bol}^{22 \mu m}$ and $L_{\rm bol}$, 
%\new{
which reveals that both estimations appear 
to be approximately consistent. 
%} 
The Spearman-rank correlation test, which assesses 
monotonic relationships, returns a coefficient of 0.48, 
which indicates that there is a moderate correlation 
between $L_{\rm bol}^{22 \mu m}$ and $L_{\rm bol}$. 
The $L_{\rm bol}^{22 \mu m}$ will be used to 
approximate $L_{\rm bol}$ for estimating the 
luminosity-to-mass ratio ($L_{\rm bol}^{22 \mu m}/M$). 
The derived $L_{\rm bol}^{22 \mu m}/M$ ranges 
from 0.03 to 2327, with a mean value of 145. 
The mean values are 6, 134 and 255 in IRDCs, 
protostars and \HII\ regions, respectively. 
The distributions of $L_{\rm bol}^{22 \mu m}/M$ 
for the three categories are shown in Figure 
\ref{fig:L22um}, which reveals that the IRDCs 
have relatively lower value than that of both 
protostars and \HII\ regions. 
We therefore use $L_{\rm bol}^{22 \mu m}/M$ as an 
approximate indicator of the evolutionary stage to 
study the SiO properties as a function of evolutionary 
stages.

%%%%%%%%%%%%%%%%%%%%%%%%%%%
%%%%%%%%%%%%%%%%%%%%%%%%%%%%%%
%DISCUSSIONS
%%%%%%%%%%%%%%%%%%%%%%%%%%%%%%
\section{Discussions}
\label{sec:discussions}
\subsection{Uncertainties of the results}
\label{uncertainties}
The uncertainty in dust continuum flux is typically 10\% 
\citep{2009A&A...504..415S,2008ApJS..175..277D}.  
The typical uncertainty in the kinematic distances is about 10\%, 
while in some cases it can be orders of magnitudes larger 
\citep{2009ApJ...700..137R}. 
Since the gas temperatures estimated from the NH$_{3}$ data 
are not available for the majority of sources, the adopted uniform 
temperature in Section \ref{column} and \ref{Mgas} could bring a 
large uncertainty into the derived parameters. Here, we adopted a 
conservative uncertainty of 10\% for temperature. The $\eta$ is 
adopted to be 100 in this study, while its standard deviation 
is 23 (corresponding to a 1$\sigma$ uncertainty of 23\%) if 
we assume that it is uniformly distributed between 70 and 150 
\citep{1990ApJ...359...42D,2003A&A...408..581V,2017ApJ...841...97S}. 
We adopted a conservative uncertainty of 28\% in $\kappa_{\nu}$ 
\citep[e.g.,][]{2017ApJ...841...97S}.  
Taking into account these uncertainties, we estimated an 
uncertainty of  44\%  and 32\% for gas mass and H$_{2}$ column 
density, respectively. However, one has to bear in mind that 
these uncertainties could be larger.

The errors of the velocity integrated intensities of the SiO 
emission are estimated with the method described in 
Section \ref{det-rate}. Note that the beam dilution effect is 
expected to affect the measured intensity. This is especially 
true in our case since the coarse spatial resolution 
(A beam of 29\arcsec\ corresponds to 0.56 pc at a distance of 
4 kpc) and a fraction of sources (31\%) are located at a 
distance of $>$ 4 kpc. The true beam-filling factor could be 
smaller than 1 in more distant  sources. The derived 
$N(\rm SiO)$ is at best regarded as the lower limit to the true 
$N(\rm SiO)$.

\subsection{Variations of SiO line widths with evolution}
\label{SiO_evol_fwzp}
The extended line wings and the broad line width (FWZP $>$ 
20 km s$^{-1}$) of the SiO emission could be due to the 
high-velocity gas driven by outflows powered by embedded 
ptotostars. Considering that such a line profile is detected 
in all types of sources, the high-velocity gas related to 
outflows should be common in various evolutionary stages, 
from the very early evolutionary stage (i.e., IRDC) to 
the more evolved (i.e., \HII\ ) stage.  On the other hand, 
we can not rule out the fact that these emission features 
contain contributions from low-velocity shocked gases 
\citep[e.g.,][]{2016A&A...595A.122L}.
We also detected 3 objects with very broad line width 
(FWZP $>$ 50 km s$^{-1}$) in the SiO emission (Figure 
\ref{fig:spec}), which is likely associated with very 
high-velocity gas with respect to the ambient gas. 
There are four sources with two distinguishable 
velocity components profile in the SiO emission 
(Figure \ref{fig:spec}). This emission feature could be 
produced by two distinctly different velocity outflows, while 
we can not rule out the other possibilities, e.g., 
episodic ejection and rotating outflows. 
These hypotheses can be tested by spatially resolving the 
kinematics of the SiO emission at higher angular resolutions.

The narrow line width (FWZP $<$ 20 km s$^{-1}$) in the 
SiO emission is detected at all three types of sources. 
This emission could be generated by several processes: 
(1) small-scale converging flows, such as the case of DR21(OH) 
\citep{2011ApJ...740L...5C} and Cygnus-X 
\citep{2014A&A...570A...1D};
(2) cloud-cloud collision in molecular clouds, such as the 
case of massive dense clump W43-MM1/MM2 
\citep{2013ApJ...775...88N,2016A&A...595A.122L}; 
(3) a population unresolved lower mass protostars 
\citep{2010MNRAS.406..187J}.  Since the 
sources have large mass reservoirs, they have a
highly potential to form multiple and/or a group of protostars.

\begin{figure*}[!tb]
	\epsscale{0.6}
	\plotone{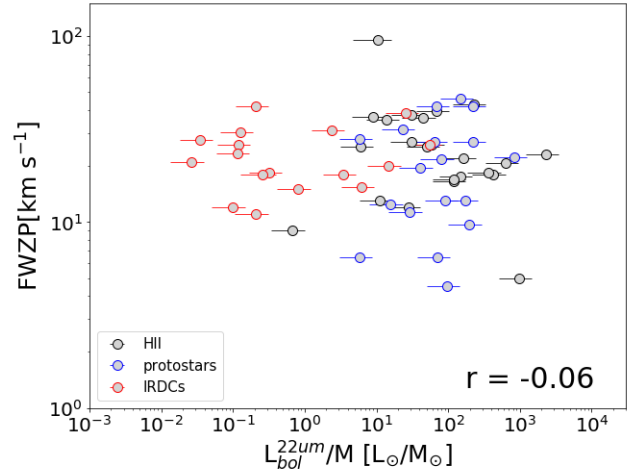}
	\caption{The FWZP of the SiO 5-4 line versus the source luminosity-to-mass ratio. 
    The grey, blue and red circles represent \HII\ regions, 
    protostars and IRDCs, respectively.
	The coefficient from the Spearman-rank correlation test 
	is presented inside the panel.}
	\label{fig:FWZP}
\end{figure*}

In shock regions, the SiO line widths are dominated by 
the non-thermal broadening. According to the 
model by \citep{2008A&A...482..809G}, the SiO intensity 
decreases as the shock passes, and so does the 
SiO observable line width. 
At the right panel of Figure \ref{fig:FWZP}, we present a 
plot of FWZP against $L_{\rm bol}^{22 \mu m}/M$. 
There is no apparent  relationship between FWZP and 
$L_{\rm bol}^{22 \mu m}/M$, and its Spearman-rank 
correlation coefficient is -0.05.  
We found no significant variations in the line width 
across the sample, which is similar to the  results 
of \cite{2016A&A...586A.149C}, who found that the FWZP of the 
SiO 2-1 line is nearly constant in different evolutionary 
stages. However, it appears to be in contrary to the results of 
\cite{2010ApJ...714.1658S} and \cite{2011A&A...526L...2L}, 
who reported that the line widths in the early evolutionary phase 
are broader than that of the more evolved evolutionary phase 
based on observations of the SiO 2-1 line toward a sample of 
massive clumps. 
Since the distance of sources that have 
$L_{\rm bol}^{22 \mu m}$ information is comparable to 
that of \cite{2010ApJ...714.1658S} and 
\cite{2011A&A...526L...2L}, it should  not be 
a significant factor for the different results. 
Several possible reasons can cause such differences: 
(1) The bias in the sample selection. 
Both \cite{2010ApJ...714.1658S} (20 sources) and 
\cite{2011A&A...526L...2L} (57 sources) studies are based 
on a relatively small sample.  
(2) This difference  could  be caused by the different 
beam sizes because our observations have a relatively large 
beam ($\sim$ 29\arcsec) than that of \cite{2010ApJ...714.1658S}  
($\sim$ 18\arcsec) and \cite{2011A&A...526L...2L} 
($\sim$ 23\arcsec), which can lead to contamination by 
nearby sources in our study. Since most of our sources 
have a mass reservoir higher than 500 M$_{\odot}$ within a 
size of 0.6 pc for a typical distance of 4 kpc, it is 
highly likely that they form a cluster or multiple sources 
that produce multiple outflows  
\citep{2015ApJ...804..141Z,2014MNRAS.439.3275W}.  
In this case, the SiO intensity may not decrease with 
the evolution of clumps thanks to the efficient SiO
replenishment from newly formed outflows/jets. 
(3) The different could also be due to limited sensitivity 
of observations for the high velocity line wings.

\subsection{Variations of SiO luminosity and 
abundance with evolution}
\label{dis:lu_Nsio}
\begin{figure*}[!tb]
	\epsscale{0.6}
	\plotone{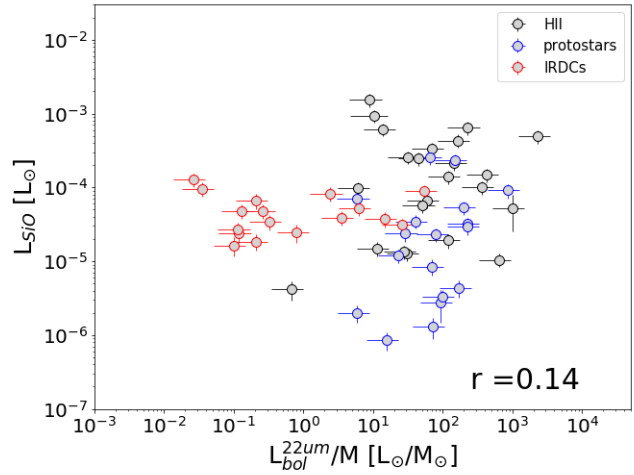}
	\caption{The SiO 5-4 luminosity versus the luminosity-to-mass ratio. 
	The grey, blue and red circle are \HII\ regions, protostars and IRDCs, respectively. The coefficient from the Spearman-rank correlation test is presented inside the panel.}
	\label{fig:L_SiO_bol}
\end{figure*}

Figure \ref{fig:L_SiO_bol} shows a plot of $L_{\rm SiO}$ against 
$L_{\rm bol}^{22 \mu m}$/M. One notes that there is 
no obvious trend between $L_{\rm SiO}$ and 
$L_{\rm bol}^{22 \mu m}$/M, with a correlation coefficient of 0.14 
between the two quantities. 
This indicates that the SiO luminosity does not change significantly 
for objects at different evolutionary stages, which is 
consistent with the results of \cite{2016A&A...586A.149C}. 
However, this appear to be in contrast with the results by 
\cite{2011A&A...526L...2L},  who found a decreasing trend in the 
SiO luminosity with the evolutionary stage,  and it was 
interpreted as either a decrease in the SiO abundance or a 
decrease in the jet/outflow energetic with the time evolution 
of star formation \citep{2011A&A...526L...2L}. 
This discrepancy could be due to the fact that the study of 
\cite{2011A&A...526L...2L} has a relatively narrow range in the 
$L_{\rm SiO}$ ($<$ 4 orders of magnitude) and $L/M$ 
($<$ 4 orders of magnitude) 
\citep[See also Figure D.2 in ][]{2016A&A...586A.149C},
while it may also be caused by the bias in the sample selection 
and/or the different beam sizes (See Section \ref{SiO_evol_fwzp}).

\begin{figure*}[!tb]
	\epsscale{1.2}
	\plotone{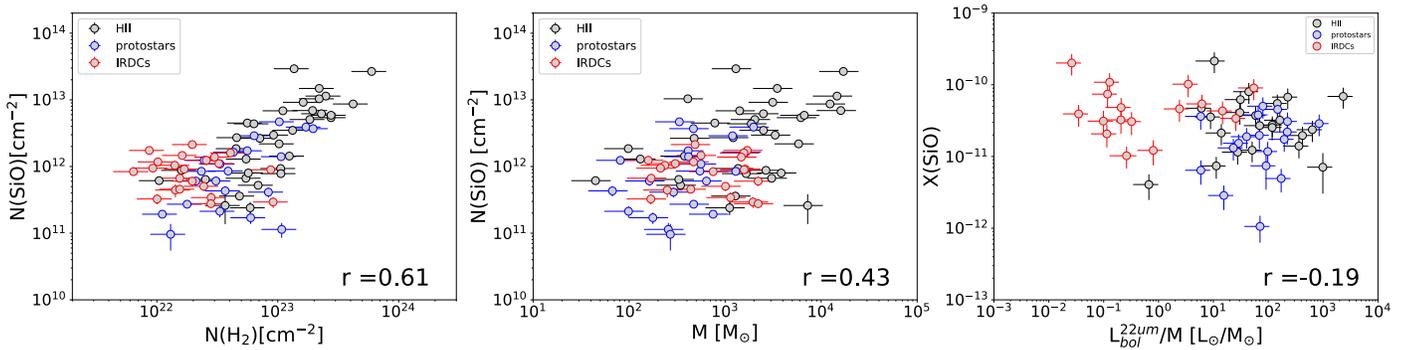}	
	\caption{\textit{Left:} SiO column density versus H$_{2}$ beam-averaged column density. 
    \textit{Middle:} The SiO column density versus gas mass.
	\textit{Right:} SiO abundance versus  luminosity-to-mass ratio. 
	The grey, blue and red circle are \HII\ regions, protostellars and IRDCs, respectively. The coefficient from the Spearman-rank correlation test is presented inside the panel.}
	\label{fig:X_SiO}
\end{figure*}

Figure \ref{fig:X_SiO} shows a comparison between $N(\rm SiO)$ 
and $N(\rm H_{2})$. The $N(\rm SiO)$ appears to increase with 
increasing $N(\rm H_{2})$, with a correlation coefficient 
of 0.61 between the two quantities. 
We also find an increasing trend between $N(\rm SiO)$ and M, 
with a correlation coefficient of 0.43 between the 
two quantities. 
These indicate that the column density of SiO  
increases as a function of dust continuum emission. 
In order to study the variation of the SiO abundance 
with evolution, we estimate the SiO abundance  
$X(\rm SiO)$ = $N(\rm SiO)$/$N(\rm N_{\rm 2})$. 
The estimated  $X(\rm SiO)$ varies between 
1.1 $\times$ 10$^{-12}$ and 2.6 $\times$ 10$^{-10}$, 
with a median value of 3.1 $\times$ 10$^{-11}$ 
and a mean value of 4.0 $\times$ 10$^{-11}$ 
(Table \ref{tab:NSiO}). 
The derived SiO abundances are comparable to 
the previous investigations 
\citep[e.g.,][]{2016A&A...586A.149C}.

The SiO abundances show no significant differences among three 
source categories, which have mean and median $X(\rm SiO)$ of around 
3 $\times$ 10$^{-11}$ (Table \ref{tab:NSiO}). 
Figure \ref{fig:X_SiO} shows a plot of  $X(\rm SiO)$  versus to 
$L_{\rm bol}^{22 \mu m}$/M.  The Spearman-rank correlation test 
returns a correlation coefficient of 0.19 with a p-value of 0.15, 
which confirms that there are not robust trends for the SiO 
abundances as a function of the evolutionary stage of massive star 
formation. A similar result has been reported in previous studies, 
i.e., \cite{2012ApJ...756...60S},  \cite{2014A&A...570A..49L} and 
\cite{2016A&A...586A.149C}. 
If these SiO emissions are due to outflows/jets, the intensity 
of the outflow/jet would not decrease as massive star formation 
proceeds. 
This speculation can be tested through higher angular resolution 
observations to spatially resolve the SiO emission within 
these massive clumps.

%%%%%%%%%%%%%%%%%%%%%%%%%%%%%%
%CONCLUSIONS
%%%%%%%%%%%%%%%%%%%%%%%%%%%%%%
\section{Summary}
\label{sec:summary}
We used the SMT 10 m and CSO 10.5 m telescopes to observe in the
SiO 5-4 line a sample of 185 and 16 massive clumps (two sources 
were observed in both telescopes), respectively. The sample
falls into three different evolutionary stages, 
IRDCs, protostars and \HII\ regions. 
Our main results are summarized as follow: 
\begin{itemize}
  \item 
For a large sample of 199 massive molecular clumps, the 
SiO 5-4 emission was detected in 102 sources, with a detection 
rate of 57\%,  37\%, and 65\% for IRDC, protostars, and \HII\ 
region, respectively. 
The high detection rate of the SiO emission across all types 
of sources indicates that the shock activities are common 
in various evolutionary stages from IRDCs to \HII\ regions. 
In addition, the presence of the SiO emission in IRDCs implies 
that there is on-going star formation in most of them. 

  \item 
Broad line widths and extended non-gaussian line 
wings are detected among all three evolutionary phases. 
These emission features are possibly associated with 
high-velocity shocked gas. We also detected narrow 
line widths with FWZP $<$ 20 $\rm km\ s^{-1}$ 
in all types of sources. They are likely associated with 
low-velocity gas with respect to the ambient gas. 
The SiO line widths show no significant differences between 
three source categories, which indicates that there is no 
robust trend between the SiO line widths and the 
evolutionary stages.

  \item 
The derived SiO luminosities ($L_{\rm SiO}$) range from 
8.5$\times$ 10$^{-7}$ to 3.7$\times$ 10$^{-3}$ $\rm L_{\odot}$ 
for this sample.
We do not find any clear trend in the $L_{\rm SiO}$ as a 
function of the evolutionary stage.

  \item
With the LTE assumption, the derived SiO total column densities 
are between 9.6 $\times 10^{10}$ and 2.9 $\times 10^{13} \rm 
cm^{-2}$. The estimated SiO abundances range from 
1.1 $\times 10^{-12}$ to 2.6 $\times 10^{-10}$, with a mean 
value of 4.0 $\times 10^{-11}$. There is no robust trend 
between the SiO abundance and the evolutionary stage, which 
suggests that the SiO abundance does not decrease with 
the advancing evolutionary stage of massive star formation 
in these massive clumps. 
\end{itemize}

High angular and high sensitivity observations will enable 
exploration in more detail of SiO outflows at various 
evolutionary stages of massive star formation.

\vspace{5mm}
\facilities{SMT, CSO.}
\software{GILDAS \citep{2005sf2a.conf..721P}}. 

\acknowledgments
We appreciate the comments from the anonymous referee 
that helped improving the paper. We are also indebted to 
the staff at Arizona Radio Observatory for their 
excellent support in the SMT observations. 
This work is supported by the National Key R$\&$D Program of 
China (No. 2017YFA0402704), the Natural Science Foundation of 
China under grants of 11590783. 
SL acknowledges support from the CfA pre-doctoral fellowship, 
the Chinese Scholarship Council, and National Natural Science 
Foundation of China grant 11629302 and U1731237. 

\bibliographystyle{aasjournal}

%%%%%%%%%%%%%%%%%
%
%\clearpage
% \bibliography{myrefs}
\bibliography{lsh-sio}

%%%%%%%%%%%%%%%%%%%%%%%%%%

%%%%%%%%%%%%%%%%%%%%%%%%%%

%%%%%%%%%%%%%%%%%%%%%%%%%%

%%%%%%%%%%%%%%%%%%%%%%%%%%
%%%%%%%%%%%%%%%%%%%%%%%%%%

%%%%%%%%%%%%%%%%%%%%%%%%%%
%%%%%%%%%%%%%%%%%%%%%%%%%%

%%%%%%%%%%%%%%%%%%%%%%%%%%
%%%%%%%%%%%%%%%%%%%%%%%%%%

%%%%%%%%%%%%%%%%%%%%%%%%

%%%%%%%%%%%%%%%%%%%%%%%%%%
%%%%%%%%%%%%%%%%%%%%%%%%%%

\clearpage

\begin{deluxetable*}{cccccc}
\tabletypesize{\scriptsize}
\tablecolumns{6}
\tablewidth{0pc}
\tablecaption{Summary of observations 
\label{tab:obs}}
\tablehead{
\colhead{Telescope--Instrument} & \colhead{Date}  & \colhead{Spectra resolution}  & \colhead{$\theta_{beam}$} &\colhead{No. of sources}
}
\startdata
SMT 10 m		&	2015	 	& 1MHz/0.25Mhz (1.38 km s$^{-1}$/0.35 km s$^{-1}$)  	& $\sim$ 28.5\arcsec 	& 88\\
SMT 10 m		&	2016	 	& 1MHz/0.25Mhz (1.38 km s$^{-1}$/0.35 km s$^{-1}$)  	& $\sim$ 28.5\arcsec  	& 97\\
CSO 10.4 m	&	2013		& 0.27MHz (0.37 km s$^{-1}$)  	& $\sim$ 27.4\arcsec 	& 16\\
\enddata
%\tablenotetext{1}{}
\end{deluxetable*}

%%%%%%%%%%%%%%%%%
%
%\input{figures.tex}
%TABLE: OBSERVATION INFORMATION
%\newpage
%{\LongTables
\setlength{\arrayrulewidth}{0.1mm}
\setlength{\tabcolsep}{2pt}
\renewcommand{\arraystretch}{1.2}

\floattable
\begin{deluxetable*}{clccccccccccccccccccccc}
\tabletypesize{\scriptsize}
\rotate
\tablecolumns{23}
\tablewidth{0pc}
\tablecaption{SiO 5-4 line parameters and source properties. \label{tbl:sio}}
\tablehead{
\colhead{No}&\colhead{Name} 	&\colhead{R.A}&\colhead{Dec} 	&\colhead{$\int T_{\rm mb}\, dv$}	&\colhead{$\varv_{lsr}$}	 &\colhead{$T_{\rm mb}$}&\colhead{FWZP}&\colhead{$N(\rm SiO)$}&\colhead{$L_{\rm SiO}$}&\colhead{$L_{\rm bol}$}&\colhead{$F\rm_{\nu}$} &\colhead{$S\rm_{\nu}$}&\colhead{$\lambda$}&\colhead{$M\rm_{gas}$}&\colhead{$N(\rm H_{2}$)} &\colhead{$S\rm_{22\mu m}$}&\colhead{$T\rm_{k}$}&\colhead{D}&\colhead{Type}&\colhead{Ref}	&\colhead{Telescope} \\
		&	 		&J2000		&J2000  				&\colhead{K km s$^{-1}$}		&\colhead{km s$^{-1}$}&\colhead{K}		&\colhead{km s$^{-1}$}&\colhead{cm$^{-2}$}&\colhead{L$_{\odot}$}		&\colhead{L$_{\odot}$}&\colhead{Jy beam$^{-1}$}&\colhead{Jy}	&\colhead{$\mu$m}	&\colhead{M$_{\odot}$}&\colhead{cm$^{-2}$}	&\colhead{Jy}&\colhead{K}&\colhead{kpc}&		&		&	     
}
\startdata
1	&	G121.30+0.66	&	00:36:47.5	&	+63:29:02	&	1.39$\pm$0.07	&	-17.2$\pm$0.1	&	0.26	&	27	&	1.85E+12	&	1.28E-05	&	1.00E+03	&	6.45	&	19.32	&	850	&	98 	&	4.52E+22	&	17.2$\pm$0.16	&	...	&	1.2	&	HII	&	1	&	SMT	\\
2	&	G123.07-6.31	&	00:52:25.2	&	+56:33:53	&	1.89$\pm$0.09	&	-30.9$\pm$0.2	&	0.25	&	22.5	&	2.52E+12	&	5.84E-05	&	...	&	...	&	...	&	...	&	...	&	...	&	2.03$\pm$0.06	&	...	&	2.2	&	HII	&	1	&	SMT	\\
3	&	G133.6945+01.2166	&	02:25:30.0	&	+62:06:20	&	0.7$\pm$0.07	&	-44$\pm$0.4	&	0.1	&	13	&	9.34E+11	&	1.79E-05	&	1.20E+04	&	15.06	&	97.04	&	850	&	1367 	&	1.06E+23	&	...			&	...	&	2	&	HII	&	3	&	SMT	\\
4	&	G133.7150+01.2155	&	02:25:40.0	&	+62:05:52	&	2.84$\pm$0.11	&	-38.9$\pm$0.1	&	0.39	&	25	&	3.91E+12	&	7.26E-05	&	1.40E+05	&	19.11	&	109.95	&	850	&	1977 	&	1.71E+23	&	...			&	...	&	2	&	protostar	&	3	&	SMT	\\
5	&	W3(OH)	&	02:27:04.6	&	+61:52:25	&	3.8$\pm$0.09	&	-47.3$\pm$0.1	&	0.49	&	16.5	&	5.07E+12	&	1.40E-04	&	8.30E+04	&	26.29	&	100.56	&	850	&	2039 	&	1.84E+23	&	349.17$\pm$1.61	&	...	&	2.4	&	UCHII	&	1	&	SMT	\\
	&	W3(OH)	&	02:27:04.6	&	+61:52:25	&	3.47$\pm$0.10	&	-48.2 $\pm$0.2	&	0.48	&	17.0	&	4.62E+12	&	1.28E-04	&	8.30E+04	&	26.29	&	100.56	&	850	&	2039 	&	1.84E+23	&	349.17 $\pm$1.61	&	...	&	2.4	&	UCHII	&	1	&	CSO	\\
6	&	G135.28+2.80	&	02:43:29.1	&	+62:56:59	&	0.46$\pm$0.05	&	-72.6$\pm$1.3	&	0.03	&	42.5	&	6.14E+11	&	1.69E-05	&	2.90E+04	&	1.49	&	2.2	&	850	&	45 	&	1.04E+22	&	...			&	...	&	2.4	&	UCHII	&	1	&	SMT	\\
7	&	05137+3919	&	05:17:13.3	&	+39:22:14	&	0.14$\pm$0.01	&	-29.4$\pm$0.5	&	0.03	&	9.7	&	1.93E+11	&	5.30E-05	&	1.70E+03	&	1.24	&	2.81	&	850	&	749 	&	1.11E+22	&	20.56$\pm$0.15	&	...	&	7.7	&	protostar	&	4	&	SMT	\\
8	&	05274+3345	&	05:30:45.6	&	+33:47:52	&	0.08$\pm$0.02	&	-3.7$\pm$0.5	&	0.03	&	6.5	&	1.14E+11	&	1.31E-06	&	5.90E+03	&	10.46	&	19.3	&	850	&	256 	&	1.08E+23	&	58.92$\pm$0.05	&	22.6	&	1.6	&	protostar	&	4	&	SMT	\\
9	&	05358+3543	&	05:39:10.4	&	+35:45:19	&	0.41$\pm$0.06	&	-16.1$\pm$1.6	&	0.03	&	42	&	6.08E+11	&	8.49E-06	&	2.90E+03	&	2.65	&	8.6	&	850	&	163 	&	3.08E+22	&	28.78$\pm$0.11	&	20.8	&	1.8	&	protostar	&	5	&	SMT	\\
10	&	S231	&	05:39:12.9	&	+35:45:54	&	1.94$\pm$0.07	&	-17.1$\pm$0.1	&	0.25	&	26	&	2.65E+12	&	6.56E-05	&	7.60E+03	&	8.25	&	20.77	&	850	&	476 	&	7.11E+22	&	44.12$\pm$0.08	&	25.7	&	2.3	&	HII	&	1	&	SMT	\\
11	&	05373+2349	&	05:40:24.4	&	+23:50:54	&	0.9$\pm$0.06	&	2.5$\pm$0.4	&	0.09	&	21.7	&	1.24E+12	&	2.30E-05	&	1.10E+03	&	2.75	&	4.5	&	850	&	81 	&	2.46E+22	&	13.45$\pm$0.16	&	...	&	2	&	protostar	&	4	&	SMT	\\
12	&	S235	&	05:40:53.3	&	+35:41:49	&	0.36$\pm$0.03	&	-17.5$\pm$0.2	&	0.08	&	10	&	5.23E+11	&	5.89E-06	&	5.30E+03	&	6.26	&	24.27	&	850	&	343 	&	6.88E+22	&	...			&	21.6	&	1.6	&	HII	&	1	&	SMT	\\
13	&	S241	&	06:03:53.6	&	+30:14:44	&	0.24$\pm$0.04	&	-7.9$\pm$0.6	&	0.04	&	12.5	&	3.20E+11	&	3.39E-05	&	...	&	...	&	...	&	...	&	...	&	...	&	1.47$\pm$0.03	&	...	&	4.7	&	HII	&	1	&	SMT	\\
14	&	S252A	&	06:08:35.4	&	+20:39:03	&	0.41$\pm$0.03	&	9.9$\pm$0.1	&	0.09	&	10.5	&	5.47E+11	&	5.89E-06	&	...	&	...	&	...	&	...	&	...	&	...	&	8.71$\pm$0.2	&	...	&	1.5	&	HII	&	1	&	SMT	\\
15	&	06056+2131	&	06:08:41.0	&	+21:31:01	&	0.3$\pm$0.05	&	2.5$\pm$0.5	&	0.05	&	13	&	4.13E+11	&	4.31E-06	&	2.40E+04	&	9.35	&	28.63	&	850	&	290 	&	8.37E+22	&	184.26$\pm$0.34	&	...	&	1.5	&	protostar	&	4	&	SMT	\\
16	&	S255	&	06:12:53.7	&	+17:59:22	&	0.97$\pm$0.07	&	7.7$\pm$0.3	&	0.12	&	20.9	&	1.29E+12	&	1.05E-05	&	4.50E+04	&	7.81	&	22.01	&	850	&	131 	&	5.47E+22	&	410.62$\pm$0.76	&	...	&	1.3	&	UCHII	&	1	&	SMT	\\
17	&	G194.9259-01.1946	&	06:13:21.0	&	+15:23:57	&	0.29$\pm$0.04	&	-5.2$\pm$0.7	&	0.04	&	15.5	&	3.87E+11	&	7.41E-06	&	1.60E+03	&	...	&	...	&	...	&	...	&	...	&	26.56$\pm$0.42	&	...	&	2	&	HII	&	3	&	SMT	\\
18	&	G5.89-0.39	&	18:00:30.3	&	-24:04:00	&	21.52$\pm$0.31	&	15.4$\pm$0.2	&	0.76	&	96	&	2.93E+13	&	9.29E-04	&	...	&	25.73	&	71.84	&	870	&	1294 	&	1.36E+23	&	16.57$\pm$4.16	&	39.7	&	2.6	&	UCHII	&	2	&	CSO	\\
19	&	G10.6-0.4	&	18:10:28.7	&	-19:55:48	&	8.54$\pm$0.11	&	-3.5$\pm$0.1	&	0.96	&	23.3	&	1.14E+13	&	1.96E-03	&	3.90E+05	&	33.1	&	107.17	&	870	&	14630 	&	2.50E+23	&	...			&	...	&	6	&	CHII	&	1	&	CSO	\\
20	&	G11.94-0.62	&	18:14:01.1	&	-18:53:24	&	0.5$\pm$0.02	&	38.8$\pm$0.4	&	0.05	&	25.5	&	6.67E+11	&	5.64E-05	&	8.30E+04	&	7.22	&	45.64	&	870	&	3053 	&	5.45E+22	&	73.22$\pm$1.28	&	...	&	4.2	&	UCHII	&	2	&	CSO	\\
21	&	W33IRS3	&	18:14:13.3	&	-17:55:40	&	4.05$\pm$0.13	&	35.8$\pm$0.1	&	0.51	&	18	&	5.40E+12	&	1.49E-04	&	3.40E+04	&	36.48	&	281.79	&	870	&	6155 	&	2.75E+23	&	3744.9	&	...	&	2.4	&	CHII	&	2	&	CSO	\\
22	&	I18151-1208MM2	&	18:17:50.0	&	-12:07:55	&	2.37$\pm$0.15	&	29.6$\pm$0.7	&	0.09	&	68.5	&	3.86E+12	&	1.36E-04	&	...	&	...	&	...	&	...	&	...	&	...	&	1.12$\pm$0.05	&	...	&	3	&	IRDC	&	6	&	SMT	\\
23	&	G015.05+00.07MM1	&	18:17:50.0	&	-15:53:38	&	0.25$\pm$0.05	&	25.4$\pm$0.2	&	0.06	&	12	&	4.44E+11	&	1.64E-05	&	...	&	0.79	&	2.69	&	870	&	249 	&	1.42E+22	&	0.02	&	16.31	&	3.2	&	IRDC	&	6	&	SMT	\\
24	&	I18151-1208MM3	&	18:17:52.0	&	-12:06:56	&	0.28$\pm$0.03	&	32.6$\pm$0.5	&	0.05	&	19	&	4.56E+11	&	1.61E-05	&	...	&	...	&	...	&	...	&	...	&	...	&	6.53$\pm$0.13	&	...	&	3	&	IRDC	&	6	&	SMT	\\
25	&	I18151-1208MM1	&	18:17:58.0	&	-12:07:27	&	0.49$\pm$0.07	&	34.2$\pm$0.5	&	0.08	&	19	&	7.98E+11	&	2.82E-05	&	2.20E+04	&	...	&	...	&	...	&	...	&	...	&	79.26$\pm$1.31	&	...	&	3	&	IRDC	&	6	&	SMT	\\
26	&	I18182-1433MM1	&	18:21:09.0	&	-14:31:57	&	1.03$\pm$0.05	&	60$\pm$0.3	&	0.11	&	26	&	1.47E+12	&	9.01E-05	&	...	&	1.47	&	7.38	&	870	&	560 	&	1.62E+22	&	18.56$\pm$0.24	&	22.67	&	3.7	&	IRDC	&	6	&	SMT	\\
27	&	I18223-1243MM4	&	18:25:07.0	&	-12:47:54	&	0.76$\pm$0.11	&	46.9$\pm$1.6	&	0.05	&	42	&	1.24E+12	&	6.65E-05	&	...	&	1.66	&	1.49	&	870	&	158 	&	2.57E+22	&	0.02	&	...	&	3.7	&	IRDC	&	6	&	SMT	\\
28	&	I18223-1243MM3	&	18:25:08.0	&	-12:45:28	&	0.53$\pm$0.06	&	45$\pm$0.3	&	0.08	&	20.5	&	8.38E+11	&	4.64E-05	&	...	&	0.44	&	7.49	&	870	&	750 	&	6.40E+21	&	...			&	18.73	&	3.7	&	IRDC	&	6	&	SMT	\\
29	&	G019.27+00.07MM2	&	18:25:53.0	&	-12:04:48	&	0.64$\pm$0.09	&	26.5$\pm$1.2	&	0.04	&	26	&	1.04E+12	&	2.36E-05	&	...	&	0.91	&	5.26	&	870	&	235 	&	1.41E+22	&	0.04	&	...	&	2.4	&	IRDC	&	6	&	SMT	\\
30	&	G019.27+00.07MM1	&	18:25:58.0	&	-12:03:59	&	1.31$\pm$0.1	&	26.5$\pm$0.6	&	0.13	&	30.5	&	2.13E+12	&	4.82E-05	&	...	&	1.27	&	10.86	&	870	&	486 	&	1.96E+22	&	0.09	&	...	&	2.4	&	IRDC	&	6	&	SMT	\\
31	&	G19.61-0.23	&	18:27:38.0	&	-11:56:28	&	4.24$\pm$0.19	&	40.8$\pm$0.3	&	0.3	&	39.5	&	6.19E+12	&	3.32E-04	&	4.60E+05	&	19.2	&	34	&	870	&	2504 	&	2.30E+23	&	118.64$\pm$1.09	&	21.4	&	3.5	&	CHIII	&	1	&	SMT	\\
32	&	G20.08-0.13	&	18:28:10.0	&	-11:28:48	&	3.37$\pm$0.18	&	42.6$\pm$0.4	&	0.25	&	36.5	&	4.49E+12	&	2.49E-04	&	1.20E+05	&	7.68	&	18.43	&	870	&	781 	&	5.60E+22	&	24.88$\pm$0.34	&	30.8	&	3.4	&	UCHII	&	1	&	SMT	\\
\enddata
\tablenotetext{}
{
$\int T_{\rm mb}\,dv$: velocity integrated intensity. 
$\varv_{lsr}$: line central velocity. 
$T_{\rm mb}$: emission peak at main beam temperature. 
FWZP: line width --  Full Widths at Zero Power. 
$N(\rm SiO)$: beam-averaged total column density of SiO.  
$L_{\rm SiO}$: beam-averaged luminosity of SiO 5-4 line. 
$L_{\rm bol}$: source bolometric luminosity. 
$F\rm_{\nu}$: peak flux density of dust continuum. 
$S\rm_{\nu}$: integrated intensity of dust continuum. 
$\lambda$: wavelength of dust continuum. 
$M\rm_{gas}$: gas mass. 
$N(\rm H_{2}$): H$_{2}$ total column density. 
$S_{22\mu m}$: source flux at WISE 22 $\mu$m. 
$T_{k}$: temperature derived from the NH$_{3}$ data. 
D: source distance. 
Type: source classification.
Ref: reference. 
\\
Reference -- (1) \cite{2003ApJS..149..375S}; 
(2) \cite{2015ApJ...802...40L}; 
(3) \cite{2015MNRAS.453..645M}; 
(4) \cite{2005ApJ...625..864Z}; 
(5) \cite{2002ApJ...566..931S}; 
(6) \cite{2008ApJ...678.1049S}. 
% (7) \cite{2011MNRAS.418.1689U}; 
% (8) \cite{2012A&A...544A.146W}.
}
\end{deluxetable*}

%%%%%  CONTINUE  %%%%%

\floattable
\begin{deluxetable*}{clccccccccccccccccccccc}
\tabletypesize{\scriptsize}
\rotate
\tablecolumns{23}
\tablewidth{0pc}
%\tablecaption{continued \label{}}
\tablehead{
\colhead{No}&\colhead{Name} 	&\colhead{R.A}&\colhead{Dec} 	&\colhead{$\int T_{\rm mb}\, dv$}	&\colhead{$\varv_{lsr}$}	 &\colhead{$T_{\rm mb}$}&\colhead{FWZP}&\colhead{$N(\rm SiO)$}&\colhead{$L_{\rm SiO}$}&\colhead{$L_{\rm bol}$}&\colhead{$F\rm_{\nu}$} &\colhead{$S\rm_{\nu}$}&\colhead{$\lambda$}&\colhead{$M\rm_{gas}$}&\colhead{$N(\rm H_{2}$)} &\colhead{$S\rm_{22\mu m}$}&\colhead{$T\rm_{k}$}&\colhead{D}&\colhead{Type}&\colhead{Ref}	&\colhead{Telescope} \\
		&	 		&J2000		&J2000  				&\colhead{K km s$^{-1}$}		&\colhead{km s$^{-1}$}&\colhead{K}		&\colhead{km s$^{-1}$}&\colhead{cm$^{-2}$}&\colhead{L$_{\odot}$}		&\colhead{L$_{\odot}$}&\colhead{Jy beam$^{-1}$}&\colhead{Jy}	&\colhead{$\mu$m}	&\colhead{M$_{\odot}$}&\colhead{cm$^{-2}$}	&\colhead{Jy}&\colhead{K}&\colhead{kpc}&		&		&	     
}
\startdata
33	&	G022.35+00.41MM2	&	18:30:24.0	&	-09:12:44	&	0.21$\pm$0.04	&	51.2$\pm$0.5	&	0.03	&	15	&	3.42E+11	&	2.48E-05	&	...	&	1.82	&	7.92	&	870	&	1137 	&	2.81E+22	&	0.41$\pm$0.02	&	...	&	4.3	&	IRDC	&	6	&	SMT	\\
34	&	G022.35+00.41MM1	&	18:30:24.0	&	-09:10:34	&	1.07$\pm$0.03	&	52.1$\pm$0.2	&	0.13	&	21	&	1.74E+12	&	1.26E-04	&	...	&	0.56	&	11.84	&	870	&	1700 	&	8.66E+21	&	0.02	&	...	&	4.3	&	IRDC	&	6	&	SMT	\\
35	&	I18306-0835MM1	&	18:33:24.0	&	-08:33:31	&	0.78$\pm$0.06	&	80.6$\pm$0.7	&	0.05	&	38.5	&	1.11E+12	&	3.11E-05	&	1.00E+04	&	2.98	&	8.8	&	870	&	302 	&	3.26E+22	&	10.32$\pm$0.31	&	22.8	&	2.5	&	IRDC	&	6	&	SMT	\\
36	&	G023.60+00.00MM3	&	18:34:10.0	&	-08:18:28	&	0.28$\pm$0.03	&	96.7$\pm$0.4	&	0.03	&	15	&	4.56E+11	&	2.72E-05	&	1.10E+04	&	1	&	3.71	&	870	&	438 	&	1.55E+22	&	...			&	...	&	3.9	&	IRDC	&	6	&	SMT	\\
37	&	G023.60+00.00MM1	&	18:34:12.0	&	-08:19:06	&	0.98$\pm$0.07	&	107.2$\pm$0.6	&	0.1	&	27.5	&	1.60E+12	&	9.52E-05	&	...	&	2.63	&	13.28	&	870	&	1569 	&	4.07E+22	&	0.03	&	...	&	3.9	&	IRDC	&	6	&	SMT	\\
38	&	G023.60+00.00MM2	&	18:34:21.0	&	-08:18:07	&	0.85$\pm$0.08	&	51.7$\pm$0.8	&	0.06	&	31	&	1.38E+12	&	8.26E-05	&	...	&	1.94	&	12.46	&	870	&	1472 	&	3.00E+22	&	1.94$\pm$0.08	&	...	&	3.9	&	IRDC	&	6	&	SMT	\\
39	&	G024.33+00.11MM1	&	18:35:08.0	&	-07:35:04	&	0.56$\pm$0.04	&	114$\pm$0.5	&	0.09	&	15.5	&	9.12E+11	&	5.17E-05	&	...	&	1.09	&	14.46	&	870	&	1622 	&	1.68E+22	&	5.78$\pm$0.11	&	...	&	3.8	&	IRDC	&	6	&	SMT	\\
40	&	G024.33+00.11MM8	&	18:35:23.0	&	-07:37:21	&	0.18$\pm$0.03	&	95.5$\pm$0.6	&	0.04	&	10.5	&	2.93E+11	&	1.66E-05	&	...	&	5.92	&	17.38	&	870	&	1949 	&	9.15E+22	&	...			&	...	&	3.8	&	IRDC	&	6	&	SMT	\\
41	&	G024.33+00.11MM3	&	18:35:28.0	&	-07:36:18	&	0.37$\pm$0.05	&	117.7$\pm$0.3	&	0.05	&	18.5	&	6.03E+11	&	3.41E-05	&	...	&	1.27	&	19.69	&	870	&	2208 	&	1.96E+22	&	0.41$\pm$0.01	&	...	&	3.8	&	IRDC	&	6	&	SMT	\\
42	&	G024.33+00.11MM2	&	18:35:34.0	&	-07:37:28	&	0.2$\pm$0.03	&	119.3$\pm$0.6	&	0.04	&	11	&	3.26E+11	&	1.85E-05	&	...	&	0.65	&	1.5	&	870	&	168 	&	1.00E+22	&	0.02	&	...	&	3.8	&	IRDC	&	6	&	SMT	\\
43	&	G024.33+00.11MM5	&	18:35:34.0	&	-07:36:42	&	0.41$\pm$0.05	&	115.5$\pm$0.3	&	0.05	&	20	&	6.68E+11	&	3.78E-05	&	...	&	1	&	1.51	&	870	&	169 	&	1.55E+22	&	1.43$\pm$0.03	&	...	&	3.8	&	IRDC	&	6	&	SMT	\\
44	&	G024.60+00.08MM2	&	18:35:36.0	&	-07:18:09	&	0.31$\pm$0.04	&	115.5$\pm$0.6	&	0.04	&	23.5	&	5.05E+11	&	2.71E-05	&	...	&	1.58	&	9.48	&	870	&	1008 	&	2.44E+22	&	0.07	&	...	&	3.7	&	IRDC	&	6	&	SMT	\\
45	&	G024.60+00.08MM1	&	18:35:40.0	&	-07:18:37	&	0.55$\pm$0.04	&	53.6$\pm$0.3	&	0.07	&	18	&	8.96E+11	&	4.81E-05	&	...	&	5.66	&	14.82	&	870	&	1576 	&	8.75E+22	&	0.25$\pm$0.04	&	...	&	3.7	&	IRDC	&	6	&	SMT	\\
46	&	G24.49-0.04	&	18:36:05.3	&	-07:31:23	&	3.41$\pm$0.17	&	110.1$\pm$0.6	&	0.21	&	47	&	4.55E+12	&	2.67E-04	&	2.10E+05	&	...	&	...	&	...	&	...	&	...	&	21.33$\pm$0.33	&	...	&	3.5	&	HII	&	1	&	SMT	\\
47	&	I18337-0743MM3	&	18:36:18.0	&	-07:41:01	&	0.72$\pm$0.05	&	55.5$\pm$0.6	&	0.06	&	31	&	1.17E+12	&	7.36E-05	&	...	&	0.66	&	3.77	&	870	&	468 	&	1.02E+22	&	...			&	...	&	4	&	IRDC	&	6	&	SMT	\\
48	&	G025.04-00.20MM1	&	18:38:10.0	&	-07:02:34	&	0.53$\pm$0.04	&	61.5$\pm$0.7	&	0.08	&	18	&	9.45E+11	&	3.91E-05	&	...	&	0.51	&	2.01	&	870	&	212 	&	9.27E+21	&	0.53$\pm$0.02	&	16.23	&	3.4	&	IRDC	&	6	&	SMT	\\
49	&	G025.04-00.20MM4	&	18:38:14.0	&	-07:03:12	&	0.17$\pm$0.02	&	65.8$\pm$0.2	&	0.07	&	6	&	2.77E+11	&	1.26E-05	&	...	&	1.81	&	24.55	&	870	&	2204 	&	2.80E+22	&	...			&	...	&	3.4	&	IRDC	&	6	&	SMT	\\
50	&	G28.86+0.07	&	18:43:46.0	&	-03:35:36	&	0.61$\pm$0.05	&	106$\pm$0.9	&	0.08	&	17.5	&	8.76E+11	&	2.13E-04	&	9.40E+04	&	3.82	&	23.13	&	870	&	2660 	&	1.60E+22	&	59.19$\pm$0.33	&	48.4	&	7.4	&	HII	&	1	&	SMT	\\
51	&	W43S	&	18:46:04.0	&	-02:39:26	&	3.21$\pm$0.15	&	97.8$\pm$0.2	&	0.35	&	23	&	4.37E+12	&	4.92E-04	&	1.40E+05	&	12.01	&	29.3	&	870	&	1869 	&	6.35E+22	&	1506.05$\pm$4.16	&	39.8	&	4.9	&	UCHII	&	1	&	SMT	\\
	&	W43S	&	18:46:04.0	&	-02:39:26	&	3.11$\pm$0.12	&	97.9 $\pm$0.3	&	0.34	&	24.0	&	4.22E+12	&	4.77E-04	&	1.40E+05	&	12.01	&	29.3	&	870	&	1869 	&	6.35E+22	&	1506.05 $\pm$4.16	&	39.8	&	4.9	&	UCHII	&	1	&	CSO	\\
52	&	G31.41+0.31	&	18:47:35.0	&	-01:12:46	&	4$\pm$0.19	&	97.1$\pm$0.3	&	0.35	&	35.5	&	5.87E+12	&	6.14E-04	&	1.40E+05	&	22.74	&	45.61	&	870	&	6672 	&	2.76E+23	&	32.23$\pm$0.33	&	21.2	&	4.9	&	UCHII	&	1	&	SMT	\\
53	&	W43Main3	&	18:47:47.0	&	-01:54:35	&	5.18$\pm$0.16	&	98.9$\pm$0.2	&	0.46	&	37	&	6.91E+12	&	1.53E-03	&	...	&	25.72	&	91.37	&	870	&	16021 	&	1.94E+23	&	25.6	&	...	&	6.8	&	HII	&	1	&	SMT	\\
54	&	G34.26+0.15	&	18:53:18.5	&	+01:14:57	&	6.5$\pm$0.21	&	59.3$\pm$0.1	&	0.93	&	19.5	&	8.67E+12	&	5.69E-04	&	...	&	55.65	&	238.32	&	870	&	12372 	&	4.20E+23	&	...			&	...	&	3.7	&	UCHII	&	2	&	CSO	\\
55	&	G35.20-0.74	&	18:58:12.7	&	+01:40:36	&	0.58$\pm$0.06	&	33$\pm$0.3	&	0.12	&	13	&	7.83E+11	&	1.48E-05	&	3.50E+04	&	12.35	&	95.8	&	870	&	1653 	&	1.06E+23	&	38.78$\pm$1.11	&	27.2	&	2	&	HII	&	1	&	CSO	\\
56	&	G45.45+0.06	&	19:14:21.3	&	+11:09:12	&	0.19$\pm$0.09	&	-51.2$\pm$0.3	&	0.08	&	5	&	2.60E+11	&	5.29E-05	&	1.20E+05	&	3.9	&	35.12	&	870	&	7239 	&	3.67E+22	&	1368.49$\pm$3.78	&	25.4	&	6.6	&	UCHII	&	2	&	CSO	\\
57	&	W51W	&	19:23:11.0	&	+14:26:38	&	0.6$\pm$0.04	&	52.1$\pm$0.3	&	0.08	&	18.5	&	7.99E+11	&	9.97E-05	&	3.70E+05	&	8.62	&	44.26	&	870	&	3832 	&	5.71E+22	&	444.47	&	33.2	&	5.1	&	HII	&	1	&	CSO	\\
58	&	W51D	&	19:23:39.9	&	+14:31:06	&	6.26$\pm$0.31	&	60.8$\pm$0.2	&	0.71	&	23	&	9.19E+12	&	1.04E-03	&	5.20E+04	&	41.67	&	61.54	&	870	&	3132 	&	1.62E+23	&	...			&	51.4	&	5.1	&	UCHII	&	2	&	CSO	\\
59	&	W51M	&	19:23:43.8	&	+14:30:29	&	19.99$\pm$0.52	&	56.6$\pm$0.1	&	1.68	&	25.5	&	2.67E+13	&	3.32E-03	&	5.10E+05	&	79.14	&	172.74	&	870	&	17038 	&	5.97E+23	&	...			&	...	&	5.1	&	CHII	&	1	&	SMT	\\
60	&	19368+2239	&	19:38:58.1	&	+22:46:32	&	0.57$\pm$0.07	&	36.6$\pm$0.5	&	0.07	&	28	&	8.47E+11	&	7.05E-05	&	1.20E+03	&	1.86	&	10.57	&	870	&	1291 	&	2.33E+22	&	3.26$\pm$0.04	&	20.7	&	4.4	&	protostar	&	4	&	SMT	\\
61	&	19388+2357	&	19:40:59.4	&	+24:04:39	&	0.2$\pm$0.02	&	34.2$\pm$0.5	&	0.03	&	11.3	&	2.72E+11	&	2.36E-05	&	2.90E+03	&	1.97	&	5.6	&	870	&	470 	&	1.78E+22	&	6.1$\pm$0.06	&	26.2	&	4.3	&	protostar	&	4	&	SMT	\\
62	&	19411+2306	&	19:43:18.1	&	+23:13:59	&	0.07$\pm$0.03	&	31.1$\pm$0.9	&	0.02	&	13	&	9.63E+10	&	2.80E-06	&	7.40E+02	&	1.35	&	8.89	&	870	&	269 	&	1.30E+22	&	32.65$\pm$0.06	&	...	&	2.5	&	protostar	&	5	&	SMT	\\
63	&	S87	&	19:46:20.5	&	+24:35:34	&	0.18$\pm$0.04	&	21$\pm$0.6	&	0.05	&	9	&	2.40E+11	&	4.15E-06	&	1.30E+04	&	8.44	&	87.59	&	850	&	1113 	&	5.91E+22	&	1.71	&	...	&	1.9	&	UCHII	&	1	&	SMT	\\
64	&	K3-50A	&	20:01:45.6	&	+33:32:42	&	1.64$\pm$0.14	&	-24.3$\pm$0.2	&	0.25	&	18	&	2.19E+12	&	6.05E-04	&	5.10E+05	&	14.64	&	28.67	&	850	&	5830 	&	1.03E+23	&	...			&	...	&	7.6	&	CHII	&	1	&	CSO	\\
65	&	20050+2720	&	20:07:06.7	&	+27:28:53	&	0.11$\pm$0.02	&	5.4$\pm$1.1	&	0.02	&	12.4	&	1.70E+11	&	8.50E-07	&	6.10E+02	&	4.6	&	22.38	&	850	&	177 	&	5.97E+22	&	18.82$\pm$0.19	&	19.3	&	1.1	&	protostar	&	4	&	SMT	\\
66	&	20056+3350	&	20:07:31.5	&	+33:59:39	&	0.25$\pm$0.03	&	8.6$\pm$0.5	&	0.03	&	14.7	&	3.44E+11	&	4.62E-06	&	...	&	...	&	...	&	...	&	...	&	...	&	23.39$\pm$0.28	&	...	&	1.7	&	protostar	&	4	&	SMT	\\
67	&	G073.0633+01.7958	&	20:08:10.0	&	+35:59:24	&	0.26$\pm$0.04	&	-0.4$\pm$0.2	&	0.12	&	4.5	&	4.31E+11	&	3.26E-06	&	3.50E+03	&	2.46	&	4.55	&	850	&	67 	&	3.68E+22	&	27.81$\pm$0.18	&	17.6	&	1.4	&	protostar	&	3	&	SMT	\\
68	&	ON1	&	20:10:09.1	&	+31:31:37	&	2.28$\pm$0.06	&	11.5$\pm$0.2	&	0.27	&	25.5	&	3.04E+12	&	5.24E-04	&	2.60E+04	&	...	&	...	&	...	&	...	&	...	&	12.42$\pm$0.08	&	...	&	6	&	UCHII	&	1	&	SMT	\\
69	&	20106+3545	&	20:12:31.3	&	+35:54:46	&	0.17$\pm$0.03	&	8$\pm$0.4	&	0.04	&	9.4	&	2.34E+11	&	2.78E-06	&	...	&	...	&	...	&	...	&	...	&	...	&	13.94$\pm$0.18	&	...	&	1.6	&	protostar	&	4	&	SMT	\\
\enddata
\tablenotetext{}{Table 1 continued}
\end{deluxetable*}

\floattable
\begin{deluxetable*}{clccccccccccccccccccccc}
\tabletypesize{\scriptsize}
\rotate
\tablecolumns{23}
\tablewidth{0pc}
%\tablecaption{continued \label{}}
\tablehead{
\colhead{No}&\colhead{Name} 	&\colhead{R.A}&\colhead{Dec} 	&\colhead{$\int T_{\rm mb}\, dv$}	&\colhead{$\varv_{lsr}$}	 &\colhead{$T_{\rm mb}$}&\colhead{FWZP}&\colhead{$N(\rm SiO)$}&\colhead{$L_{\rm SiO}$}&\colhead{$L_{\rm bol}$}&\colhead{$F\rm_{\nu}$} &\colhead{$S\rm_{\nu}$}&\colhead{$\lambda$}&\colhead{$M\rm_{gas}$}&\colhead{$N(\rm H_{2}$)} &\colhead{$S\rm_{22\mu m}$}&\colhead{$T\rm_{k}$}&\colhead{D}&\colhead{Type}&\colhead{Ref}	&\colhead{Telescope} \\
		&	 		&J2000		&J2000  				&\colhead{K km s$^{-1}$}		&\colhead{km s$^{-1}$}&\colhead{K}		&\colhead{km s$^{-1}$}&\colhead{cm$^{-2}$}&\colhead{L$_{\odot}$}		&\colhead{L$_{\odot}$}&\colhead{Jy beam$^{-1}$}&\colhead{Jy}	&\colhead{$\mu$m}	&\colhead{M$_{\odot}$}&\colhead{cm$^{-2}$}	&\colhead{Jy}&\colhead{K}&\colhead{kpc}&		&		&	     
}
\startdata
70	&	20126+4104	&	20:14:26.0	&	+41:13:32	&	2.05$\pm$0.15	&	-3.7$\pm$0.8	&	0.11	&	46	&	2.89E+12	&	2.31E-04	&	4.00E+03	&	6.38	&	13.48	&	850	&	1194 	&	6.37E+22	&	83.68$\pm$0.15	&	23.1	&	4.2	&	protostar	&	4	&	SMT	\\
71	&	20188+3928	&	20:20:39.3	&	+39:37:52	&	0.71$\pm$0.05	&	0.7$\pm$0.3	&	0.07	&	27.5	&	9.77E+11	&	1.81E-05	&	2.20E+03	&	...	&	...	&	...	&	...	&	...	&	187.34$\pm$1.38	&	...	&	2	&	protostar	&	4	&	SMT	\\
72	&	ON2S	&	20:21:41.0	&	+37:25:29	&	1.93$\pm$0.17	&	-2.3$\pm$0.8	&	0.11	&	39	&	2.71E+12	&	3.73E-04	&	2.90E+02	&	10.86	&	18.95	&	850	&	1212 	&	4.57E+22	&	...			&	45.1	&	5.5	&	HII	&	1	&	SMT	\\
73	&	ON2N	&	20:21:43.9	&	+37:26:39	&	2.22$\pm$0.12	&	-0.2$\pm$0.3	&	0.22	&	22	&	2.96E+12	&	4.29E-04	&	1.50E+04	&	13.17	&	31.14	&	850	&	3316 	&	9.23E+22	&	148.4$\pm$0.14	&	...	&	5.5	&	CHII	&	1	&	SMT	\\
74	&	S106	&	20:27:25.7	&	+37:22:52	&	0.27$\pm$0.03	&	-5.9$\pm$0.6	&	0.04	&	16	&	3.60E+11	&	2.90E-05	&	...	&	6.89	&	21.57	&	850	&	1277 	&	4.83E+22	&	...		&		...	&	4.1	&	UCHII	&	1	&	SMT	\\
75	&	20286+4105	&	20:30:27.9	&	+41:15:48	&	0.29$\pm$0.05	&	-2.4$\pm$0.7	&	0.04	&	17.2	&	3.99E+11	&	2.54E-05	&	1.10E+03	&	...	&	...	&	...	&	...	&	...	&	...			&	...	&	3.7	&	protostar	&	4	&	SMT	\\
76	&	20293+3952	&	20:31:10.7	&	+40:03:10	&	1.14$\pm$0.09	&	12$\pm$0.9	&	0.08	&	42	&	1.72E+12	&	3.21E-05	&	3.50E+02	&	4.58	&	15.24	&	850	&	412 	&	5.59E+22	&	174.52$\pm$0.64	&	20.1	&	2.1	&	protostar	&	5	&	SMT	\\
77	&	G083.7962+03.3058	&	20:33:48.0	&	+45:40:54	&	1.08$\pm$0.05	&	-9.3$\pm$0.6	&	0.11	&	19.5	&	1.44E+12	&	1.35E-05	&	1.20E+04	&	...	&	...	&	...	&	...	&	...	&	...			&	...	&	1.4	&	HII	&	3	&	SMT	\\
78	&	20333+4102	&	20:35:09.5	&	+41:13:18	&	0.27$\pm$0.04	&	4.6$\pm$0.5	&	0.04	&	11.5	&	3.71E+11	&	4.42E-06	&	...	&	...	&	...	&	...	&	...	&	...	&	25.04$\pm$0.28	&	...	&	1.6	&	protostar	&	4	&	SMT	\\
79	&	G080.8645+00.4197	&	20:36:52.0	&	+41:36:24	&	0.29$\pm$0.04	&	-3.5$\pm$0.6	&	0.06	&	10.5	&	3.87E+11	&	3.63E-06	&	4.80E+03	&	...	&	...	&	...	&	...	&	...	&	174.84$\pm$0.64	&	...	&	1.4	&	HII	&	3	&	SMT	\\
80	&	G080.8624+00.3827	&	20:37:00.0	&	+41:34:55	&	0.16$\pm$0.03	&	-0.3$\pm$0.3	&	0.04	&	6.5	&	2.13E+11	&	2.00E-06	&	2.00E+03	&	5.17	&	15.58	&	850	&	98 	&	3.32E+22	&	2.47$\pm$0.05	&	32.1	&	1.4	&	protostar	&	3	&	SMT	\\
81	&	G081.8652+00.7800	&	20:38:35.0	&	+42:37:13	&	2.77$\pm$0.1	&	10.9$\pm$0.2	&	0.33	&	19.5	&	3.70E+12	&	3.47E-05	&	4.90E+03	&	33.58	&	81.35	&	850	&	467 	&	1.96E+23	&	79.99$\pm$0.44	&	34.6	&	1.4	&	protostar	&	3	&	SMT	\\
82	&	G081.8789+00.7822	&	20:38:37.0	&	+42:37:58	&	0.7$\pm$0.1	&	10.5$\pm$0.6	&	0.12	&	15	&	9.34E+11	&	8.77E-06	&	1.10E+04	&	...	&	...	&	...	&	...	&	...	&	917.57$\pm$4.23	&	...	&	1.4	&	HII	&	3	&	SMT	\\
83	&	G081.7131+00.5792	&	20:38:57.0	&	+42:22:40	&	0.48$\pm$0.07	&	1.8$\pm$0.9	&	0.05	&	18.5	&	6.60E+11	&	6.01E-06	&	3.60E+03	&	...	&	...	&	...	&	...	&	...	&	109.31$\pm$0.6	&	...	&	1.4	&	protostar	&	3	&	SMT	\\
84	&	DR21S	&	20:39:00.8	&	+42:19:29	&	11.12$\pm$0.23	&	-0.9$\pm$0.2	&	0.65	&	43	&	1.48E+13	&	6.39E-04	&	6.90E+02	&	31.42	&	110.09	&	850	&	3488 	&	2.20E+23	&	731.54	&	...	&	3	&	UCHII	&	1	&	CSO	\\
85	&	G081.7220+00.5699	&	20:39:01.0	&	+42:22:50	&	7.79$\pm$0.15	&	-2.9$\pm$0.1	&	0.86	&	25.5	&	1.04E+13	&	9.75E-05	&	1.90E+03	&	34.15	&	64.35	&	850	&	408 	&	2.20E+23	&	10.47$\pm$0.25	&	32	&	1.4	&	HII	&	3	&	SMT	\\
86	&	G081.7522+00.5906	&	20:39:01.0	&	+42:24:59	&	0.95$\pm$0.08	&	-2.3$\pm$0.7	&	0.08	&	31.5	&	1.40E+12	&	1.19E-05	&	9.00E+03	&	9.24	&	36.55	&	850	&	415 	&	1.07E+23	&	40.72$\pm$1.31	&	20.9	&	1.4	&	protostar	&	3	&	SMT	\\
87	&	G081.7133+00.5589	&	20:39:02.0	&	+42:21:58	&	1.07$\pm$0.08	&	-3.2$\pm$0.2	&	0.2	&	12	&	1.43E+12	&	1.34E-05	&	9.10E+03	&	16.7	&	51.38	&	850	&	378 	&	1.25E+23	&	44.45$\pm$0.66	&	28.6	&	1.4	&	HII	&	3	&	SMT	\\
88	&	G081.7624+00.5916	&	20:39:03.0	&	+42:25:29	&	2.98$\pm$0.16	&	3.5$\pm$0.8	&	0.17	&	37.5	&	4.68E+12	&	3.73E-05	&	2.00E+03	&	7.69	&	25.27	&	850	&	333 	&	1.03E+23	&	...			&	18.9	&	1.4	&	protostar	&	3	&	SMT	\\
89	&	G97.53+3.19	&	21:32:11.4	&	+55:53:55	&	0.44$\pm$0.04	&	-71.4$\pm$0.5	&	0.04	&	25.5	&	5.87E+11	&	1.34E-04	&	1.00E+04	&	...	&	...	&	...	&	...	&	...	&	7.01$\pm$0.09	&	...	&	6.9	&	HII	&	1	&	SMT	\\
90	&	G094.4637-00.8043	&	21:35:09.0	&	+50:53:09	&	1.35$\pm$0.1	&	-47.2$\pm$0.5	&	0.17	&	19	&	1.86E+12	&	2.16E-04	&	2.10E+04	&	...	&	...	&	...	&	...	&	...	&	39.76$\pm$0.04	&	...	&	5	&	protostar	&	3	&	SMT	\\
91	&	21391+5802	&	21:40:42.4	&	+58:16:10	&	3.18$\pm$0.19	&	3.4$\pm$0.7	&	0.14	&	57	&	4.37E+12	&	1.30E-05	&	...	&	...	&	...	&	...	&	...	&	...	&	1.93$\pm$0.04	&	...	&	0.8	&	protostar	&	4	&	SMT	\\
92	&	BFS11-B	&	21:43:06.7	&	+66:07:04	&	0.23$\pm$0.03	&	-8.5$\pm$0.5	&	0.05	&	14.5	&	3.07E+11	&	9.18E-06	&	...	&	...	&	...	&	...	&	...	&	...	&	41.48$\pm$0.08	&	...	&	2.5	&	HII	&	1	&	SMT	\\
93	&	G105.5072+00.2294	&	22:32:23.0	&	+58:18:58	&	0.14$\pm$0.02	&	-52.1$\pm$0.3	&	0.05	&	36	&	1.93E+11	&	1.89E-05	&	7.00E+03	&	...	&	...	&	...	&	...	&	...	&	15.23$\pm$0.08	&	...	&	4.6	&	protostar	&	3	&	SMT	\\
94	&	22506+5944	&	22:52:38.6	&	+60:00:56	&	1.22$\pm$0.09	&	-48.3$\pm$0.4	&	0.11	&	27	&	1.64E+12	&	2.53E-04	&	4.20E+04	&	5.71	&	13.32	&	850	&	1638 	&	4.30E+22	&	27.46$\pm$0.2	&	28.4	&	5.7	&	protostar	&	4	&	SMT	\\
95	&	23033+5951	&	23:05:25.5	&	+60:08:06	&	0.41$\pm$0.06	&	-51.2$\pm$0.7	&	0.05	&	22.5	&	5.64E+11	&	3.21E-05	&	...	&	...	&	...	&	...	&	...	&	...	&	9.67$\pm$0.09	&	...	&	3.5	&	protostar	&	5	&	SMT	\\
96	&	NGC7538	&	23:13:44.8	&	+61:26:50	&	5.11$\pm$0.1	&	-55.2$\pm$0.1	&	0.54	&	37.5	&	6.85E+12	&	2.56E-04	&	4.60E+03	&	14.64	&	47.72	&	850	&	1416 	&	1.10E+23	&	46.71$\pm$0.13	&	28.4	&	2.8	&	UCHII	&	1	&	SMT	\\
97	&	NGC7538A	&	23:13:45.6	&	+61:28:18	&	2.62$\pm$0.07	&	-57.2$\pm$0.1	&	0.36	&	18.5	&	3.49E+12	&	2.05E-04	&	...	&	19.59	&	60.16	&	850	&	2507 	&	1.33E+23	&	...			&	30.8	&	3.5	&	UCHII	&	2	&	CSO	\\
98	&	G111.5671+00.7517	&	23:14:01.0	&	+61:27:19	&	0.63$\pm$0.06	&	-58.8$\pm$0.9	&	0.06	&	27	&	8.55E+11	&	2.93E-05	&	2.30E+04	&	4.7	&	18.03	&	850	&	546 	&	3.88E+22	&	138.62$\pm$0.64	&	26.5	&	2.7	&	protostar	&	3	&	SMT	\\
99	&	S157	&	23:16:04.4	&	+60:01:41	&	0.48$\pm$0.05	&	-42.7$\pm$0.4	&	0.06	&	17	&	6.40E+11	&	1.92E-05	&	1.30E+04	&	3.63	&	15.08	&	850	&	329 	&	2.52E+22	&	51.65$\pm$0.86	&	30.2	&	2.5	&	CHII	&	1	&	SMT	\\
100	&	23139+5939	&	23:16:08.7	&	+59:55:18	&	0.81$\pm$0.07	&	-45.5$\pm$0.4	&	0.09	&	17	&	1.09E+12	&	1.19E-04	&	1.10E+04	&	4.28	&	6.21	&	850	&	560 	&	3.33E+22	&	...			&	27.7	&	4.8	&	protostar	&	5	&	SMT	\\
101	&	23151+5912	&	23:17:21.0	&	+59:28:49	&	0.44$\pm$0.04	&	-54.5$\pm$0.3	&	0.07	&	22.3	&	6.05E+11	&	9.13E-05	&	1.10E+03	&	2.35	&	4.37	&	850	&	638 	&	2.10E+22	&	138.5$\pm$0.26	&	...	&	5.7	&	protostar	&	5	&	SMT	\\
102	&	23385+6053	&	23:40:53.2	&	+61:10:21	&	0.36$\pm$0.04	&	-49.3$\pm$0.7	&	0.06	&	12.5	&	4.95E+11	&	1.10E-04	&	...	&	...	&	...	&	...	&	...	&	...	&	2.08$\pm$0.06	&	...	&	6.9	&	protostar	&	4	&	SMT	\\
\enddata
\tablenotetext{}{Table 1 continued}
\end{deluxetable*}

%TABLE: OBSERVATION INFORMATION
%\clearpage

\setlength{\arrayrulewidth}{0.1mm}
\setlength{\tabcolsep}{2pt}
\renewcommand{\arraystretch}{1.2}

\floattable
\begin{deluxetable*}{clccccccc}
\tabletypesize{\scriptsize}
\tablecolumns{9}
\tablewidth{0pc}
\tablecaption{The parameters of sources without SiO 5-4 detection.  \label{tbl:no-sio}}
\tablehead{
\colhead{No}&\colhead{Name} 	&\colhead{R.A. (J2000)}&\colhead{Dec. (J2000)}	&\colhead{$\varv_{lsr}$}	&\colhead{Distance} &\colhead{Type}&\colhead{Ref} &\colhead{Telescope}\\
		&		&				    hh:mm:ss&dd:mm:ss		&\colhead{kms$^{-1}$}	&\colhead{kpc} 		&			& &
}
\startdata
1	&	00117+6412	&	00:14:27.7	&	+64:28:46	&	-36.3	&	1.8	&	protostar	&	4	&	SMT	\\
2	&	00420+5530	&	00:44:57.6	&	+55:47:18	&	-51.2	&	7.7	&	protostar	&	4	&	SMT	\\
3	&	G125.6045+02.1038	&	01:16:36.0	&	+64:50:38	&	-53.7	&	4.1	&	UCHII	&	3	&	SMT	\\
4	&	G125.7795+01.7285	&	01:17:53.0	&	+64:27:14	&	-64.5	&	5.2	&	protostar	&	3	&	SMT	\\
5	&	G132.1570-00.7241	&	02:08:05.0	&	+60:46:02	&	-55.6	&	4.5	&	HII	&	3	&	SMT	\\
6	&	G134.2792+00.8561	&	02:29:01.0	&	+61:33:30	&	-51.5	&	2	&	protostar	&	3	&	SMT	\\
7	&	G136.3833+02.2666	&	02:50:08.0	&	+61:59:52	&	-42.4	&	3.3	&	protostar	&	3	&	SMT	\\
8	&	G138.2957+01.5552	&	03:01:31.0	&	+60:29:13	&	-38	&	2.9	&	protostar	&	3	&	SMT	\\
9	&	G139.9091+00.1969	&	03:07:24.0	&	+58:30:43	&	-39.5	&	3.2	&	UCHII	&	3	&	SMT	\\
10	&	G142.2446+01.4299	&	03:27:31.0	&	+58:19:21	&	-46.7	&	4.2	&	HII	&	3	&	SMT	\\
11	&	G141.9996+01.8202	&	03:27:38.0	&	+58:47:00	&	-13.9	&	0.8	&	protostar	&	3	&	SMT	\\
12	&	04579+4703	&	05:01:39.7	&	+47:07:23	&	-16.5	&	2.5	&	protostar	&	4	&	SMT	\\
13	&	05168+3634	&	05:20:16.2	&	+36:37:21	&	-15.1	&	6.1	&	protostar	&	4	&	SMT	\\
14	&	05345+3157	&	05:37:47.8	&	+31:59:24	&	-18.4	&	1.8	&	protostar	&	4	&	SMT	\\
15	&	05490+2658	&	05:52:12.9	&	+26:59:33	&	0.8	&	2.1	&	protostar	&	5	&	SMT	\\
16	&	G192.5843-00.0417	&	06:12:53.0	&	+18:00:26	&	8.8	&	2	&	HII	&	3	&	SMT	\\
17	&	G192.6005-00.0479	&	06:12:54.0	&	+17:59:23	&	7.4	&	2	&	protostar	&	3	&	SMT	\\
18	&	G194.9349-01.2224	&	06:13:16.0	&	+15:22:43	&	15.9	&	2	&	protostar	&	3	&	SMT	\\
19	&	G207.2654-01.8080	&	06:34:37.0	&	+04:12:44	&	12.6	&	1	&	UCHII	&	3	&	SMT	\\
20	&	G212.0641-00.7395	&	06:47:13.0	&	+00:26:06	&	45	&	4.7	&	protostar	&	3	&	SMT	\\
21	&	RCW142	&	17:50:15.1	&	+28:54:32	&	17	&	2	&	UCHII	&	1	&	SMT	\\
22	&	G9.62+0.10	&	18:06:14.8	&	+20:31:40	&	2.5	&	7	&	UCHII	&	1	&	SMT	\\
23	&	G12.42+0.50	&	18:10:51.8	&	-17:55:56	&	17.7	&	2.1	&	UCHII	&	1	&	SMT	\\
24	&	G12.89+0.49	&	18:11:51.6	&	-17:52:24	&	33.8	&	3.5	&	HII	&	1	&	SMT	\\
25	&	G12.2-0.1	&	18:12:39.7	&	-18:24:20	&	24.4	&	3.7	&	CHII	&	1	&	SMT	\\
26	&	G13.87+0.28	&	18:14:35.0	&	-16:15:37	&	49	&	4.4	&	HII	&	1	&	SMT	\\
27	&	W33A	&	18:14:39.3	&	+17:52:11	&	36.6	&	4.5	&	HII	&	1	&	SMT	\\
28	&	G015.31-00.16MM2	&	18:18:50.0	&	-15:43:19	&	31.1	&	3.2	&	IRDC	&	6	&	SMT	\\
29	&	I18182-1433MM2	&	18:21:15.0	&	-14:33:06	&	41.1	&	3.6	&	IRDC	&	6	&	SMT	\\
30	&	G023.60+00.00MM4	&	18:25:07.0	&	-08:18:21	&	53.8	&	3.9	&	IRDC	&	6	&	SMT	\\
31	&	I18223-1243MM2	&	18:25:09.0	&	-12:44:15	&	45.3	&	3.7	&	IRDC	&	6	&	SMT	\\
32	&	I18223-1243MM1	&	18:25:10.0	&	-12:42:26	&	45	&	3.7	&	IRDC	&	6	&	SMT	\\
33	&	G022.35+00.41MM2	&	18:30:24.0	&	-09:12:44	&	60.2	&	4.3	&	IRDC	&	6	&	SMT	\\
34	&	I18306-0835MM2	&	18:33:17.0	&	-08:33:26	&	76.7	&	2.5	&	IRDC	&	6	&	SMT	\\
35	&	I18306-0835MM3	&	18:33:32.0	&	-08:32:29	&	53.9	&	2.5	&	IRDC	&	6	&	SMT	\\
36	&	G023.60+00.00MM7	&	18:34:21.0	&	-08:17:11	&	54	&	3.9	&	IRDC	&	6	&	SMT	\\
37	&	G23.95+0.16	&	18:34:23.8	&	+07:54:53	&	79.9	&	5.8	&	HII	&	1	&	SMT	\\
38	&	G024.08+00.04MM2	&	18:34:51.0	&	-07:45:32	&	52.1	&	3.8	&	IRDC	&	6	&	SMT	\\
39	&	G024.08+00.04MM1	&	18:34:57.0	&	-07:43:26	&	113.6	&	3.8	&	IRDC	&	6	&	SMT	\\
40	&	G024.08+00.04MM3	&	18:35:02.0	&	-07:45:25	&	51.8	&	3.8	&	IRDC	&	6	&	SMT	\\
41	&	G024.08+00.04MM4	&	18:35:03.0	&	-07:45:56	&	51.9	&	3.8	&	IRDC	&	6	&	SMT	\\
42	&	G024.33+00.11MM11	&	18:35:05.0	&	-07:35:58	&	112.6	&	3.8	&	IRDC	&	6	&	SMT	\\
43	&	G024.33+00.11MM6	&	18:35:08.0	&	-07:34:33	&	114.4	&	3.8	&	IRDC	&	6	&	SMT	\\
44	&	G024.33+00.11MM4	&	18:35:19.0	&	-07:37:17	&	115.1	&	3.8	&	IRDC	&	6	&	SMT	\\
45	&	G024.33+00.11MM9	&	18:35:26.0	&	-07:36:56	&	119.7	&	3.8	&	IRDC	&	6	&	SMT	\\
46	&	I18337-0743MM2	&	18:36:28.0	&	-07:40:28	&	58.6	&	4	&	IRDC	&	6	&	SMT	\\
47	&	G025.04-00.20MM2	&	18:38:18.0	&	-07:02:51	&	63.8	&	3.4	&	IRDC	&	6	&	SMT	\\
48	&	G35.20-1.74	&	19:01:47.0	&	+01:13:08	&	43.6	&	3.5	&	UCHII	&	2	&	CSO	\\
\enddata
\tablenotetext{}{
Reference. -- (1) \cite{2003ApJS..149..375S}; (2) \cite{2015ApJ...802...40L}; (3) \cite{2015MNRAS.453..645M}; (4) \cite{2005ApJ...625..864Z}; (5) \cite{2002ApJ...566..931S}; (6) \cite{2008ApJ...678.1049S}. }
\end{deluxetable*}

\floattable
\begin{deluxetable*}{clccccccc}
\tabletypesize{\scriptsize}
\tablecolumns{9}
\tablewidth{0pc}
%\tablecaption{Sources of no detected SiO emission %\label{tbl:no-sio}}
\tablehead{
\colhead{No}&\colhead{Name} 	&\colhead{R.A. (J2000)}&\colhead{Dec. (J2000)}	&\colhead{$\varv_{lsr}$}	&\colhead{Distance} &\colhead{Type}&\colhead{Ref} &\colhead{Telescope}\\
		&		&				    hh:mm:ss&dd:mm:ss		&\colhead{kms$^{-1}$}	&\colhead{kpc} 		&			& &
}
\startdata
49	&	19282+1814	&	19:30:28.1	&	+18:20:53	&	24.1	&	2.1	&	protostar	&	4	&	SMT	\\
50	&	19374+2352	&	19:39:33.2	&	+23:59:55	&	36.9	&	4.3	&	protostar	&	4	&	SMT	\\
51	&	19403+2258	&	19:42:27.2	&	+23:05:12	&	26.7	&	2.4	&	protostar	&	5	&	SMT	\\
52	&	G59.78+0.06	&	19:43:11.6	&	+23:43:54	&	22.4	&	2.2	&	UCHII	&	1	&	SMT	\\
53	&	19413+2332	&	19:43:28.9	&	+23:40:04	&	20.8	&	1.8	&	protostar	&	5	&	SMT	\\
54	&	G61.48+0.09	&	19:46:47.3	&	+25:12:45	&	21.9	&	5.4	&	UCHII	&	2	&	CSO	\\
55	&	19471+2641	&	19:49:09.9	&	+26:48:52	&	21	&	2.4	&	protostar	&	5	&	SMT	\\
56	&	20051+3435	&	20:07:03.8	&	+34:44:35	&	11.6	&	1.6	&	protostar	&	5	&	SMT	\\
57	&	20062+3550	&	20:08:09.8	&	+35:59:20	&	0.6	&	4.9	&	protostar	&	4	&	SMT	\\
58	&	20081+2720	&	20:10:11.5	&	+27:29:06	&	5.7	&	0.7	&	protostar	&	5	&	SMT	\\
59	&	20099+3640	&	20:11:46.4	&	+36:49:37	&	-36.4	&	8.7	&	protostar	&	4	&	SMT	\\
60	&	20216+4107	&	20:23:23.8	&	+41:17:40	&	-20	&	1.7	&	protostar	&	5	&	SMT	\\
61	&	20217+3947	&	20:23:31.7	&	+39:57:23	&	-0.9	&	3.7	&	protostar	&	4	&	SMT	\\
62	&	20220+3728	&	20:23:55.7	&	+37:38:10	&	-2.7	&	4.5	&	protostar	&	4	&	SMT	\\
63	&	G077.9550+00.0058	&	20:29:31.0	&	+39:01:20	&	-4.2	&	1.4	&	HII	&	3	&	SMT	\\
64	&	20278+3521	&	20:29:46.9	&	+35:31:39	&	-4.5	&	5	&	protostar	&	4	&	SMT	\\
65	&	G079.8749+01.1821	&	20:30:27.0	&	+41:15:59	&	-3.8	&	1.4	&	HII	&	3	&	SMT	\\
66	&	G083.0936+03.2724	&	20:31:35.0	&	+45:05:45	&	-3	&	1.4	&	HII	&	3	&	SMT	\\
67	&	G083.7071+03.2817	&	20:33:36.0	&	+45:35:44	&	-3.2	&	1.4	&	protostar	&	3	&	SMT	\\
68	&	20319+3958	&	20:33:49.3	&	+40:08:45	&	8.8	&	1.6	&	protostar	&	5	&	SMT	\\
69	&	20332+4124	&	20:34:59.9	&	+41:34:49	&	-20	&	3.9	&	protostar	&	5	&	SMT	\\
70	&	20343+4129	&	20:36:07.1	&	+41:40:01	&	11.5	&	1.4	&	protostar	&	5	&	SMT	\\
71	&	G080.9383-00.1268	&	20:39:25.0	&	+41:20:01	&	-1.2	&	1.4	&	HII	&	3	&	SMT	\\
72	&	20444+4629	&	20:46:08.3	&	+46:40:41	&	-4.1	&	2.4	&	protostar	&	4	&	SMT	\\
73	&	G085.4102+00.0032	&	20:54:14.0	&	+44:54:04	&	-36.5	&	5.5	&	protostar	&	3	&	SMT	\\
74	&	21078+5211	&	21:09:25.2	&	+52:23:44	&	-6.1	&	1.5	&	protostar	&	4	&	SMT	\\
75	&	G094.3228-00.1671	&	21:31:45.0	&	+51:15:35	&	-38.5	&	4.4	&	protostar	&	3	&	SMT	\\
76	&	21307+5049	&	21:32:31.5	&	+51:02:22	&	-46.7	&	4.9	&	protostar	&	4	&	SMT	\\
77	&	G094.6028-01.7966	&	21:39:58.0	&	+50:14:20	&	-43.9	&	4.9	&	protostar	&	3	&	SMT	\\
78	&	21519+5613	&	21:53:39.2	&	+56:27:46	&	-63.2	&	7.3	&	protostar	&	4	&	SMT	\\
79	&	22134+5834	&	22:15:09.1	&	+58:49:09	&	-18.3	&	2.6	&	protostar	&	5	&	SMT	\\
80	&	22172+5549	&	22:19:09.0	&	+56:04:45	&	-43.8	&	2.9	&	protostar	&	4	&	SMT	\\
81	&	22198+6336	&	22:21:27.6	&	+63:51:42	&	-11.1	&	1.3	&	protostar	&	4	&	SMT	\\
82	&	22305+5803	&	22:32:24.3	&	+58:18:58	&	-52.1	&	5.4	&	protostar	&	4	&	SMT	\\
83	&	G105.6270+00.3388	&	22:32:45.0	&	+58:28:18	&	-52.1	&	4.5	&	HII	&	3	&	SMT	\\
84	&	G107.6823-02.2423	&	22:55:29.0	&	+58:09:24	&	-55.1	&	4.7	&	protostar	&	3	&	SMT	\\
85	&	22551+6221	&	22:57:05.2	&	+62:37:44	&	-13.4	&	0.7	&	protostar	&	5	&	SMT	\\
86	&	G108.7575-00.9863	&	22:58:47.0	&	+58:45:01	&	-50.8	&	4.3	&	protostar	&	3	&	SMT	\\
87	&	G109.0775-00.3524	&	22:58:58.0	&	+59:27:36	&	-48.3	&	4	&	protostar	&	3	&	SMT	\\
88	&	G109.0974-00.3458	&	22:59:05.0	&	+59:28:23	&	-46.7	&	3.8	&	HII	&	3	&	SMT	\\
89	&	G108.4714-02.8176	&	23:02:32.0	&	+56:57:51	&	-53.9	&	4.5	&	protostar	&	3	&	SMT	\\
90	&	23026+5948	&	23:04:45.7	&	+60:04:35	&	-51.1	&	5.8	&	protostar	&	4	&	SMT	\\
91	&	G110.1082+00.0473	&	23:05:10.0	&	+60:14:47	&	-50.2	&	4.3	&	HII	&	3	&	SMT	\\
92	&	G111.5234+00.8004	&	23:13:32.0	&	+61:29:06	&	-58	&	2.7	&	protostar	&	3	&	SMT	\\
93	&	G111.5851+00.7976	&	23:14:01.0	&	+61:30:17	&	-56.5	&	2.7	&	protostar	&	3	&	SMT	\\
94	&	23140+6121	&	23:16:11.7	&	+61:37:45	&	-51.5	&	6.4	&	protostar	&	4	&	SMT	\\
95	&	G114.0835+02.8568	&	23:28:27.0	&	+64:17:38	&	-53.2	&	4.2	&	protostar	&	3	&	SMT	\\
96	&	23314+6033	&	23:33:44.3	&	+60:50:30	&	-45.4	&	2.8	&	protostar	&	4	&	SMT	\\
97	&	23545+6508	&	23:57:05.2	&	+65:25:11	&	-18.4	&	0.8	&	protostar	&	5	&	SMT	\\
\enddata
\tablenotetext{}{Table 2 continued}
\end{deluxetable*}

%%%%%%%%%%%%%%%%%

%%%%%%%%%%%%%%%%%
%
\begin{deluxetable*}{cccccccc}
\tabletypesize{\scriptsize}
\tablecolumns{7}
\tablewidth{0pc}
\tablecaption{The SiO 5-4 line widths.  
\label{tab:vel}}
\tablehead{
\colhead{Source groups} & \colhead{total} & \colhead{FWZP $\leqslant$ 10} & \colhead{ 10 $<$ FWZP $\leqslant$ 20}  & \colhead{20 $<$ FWZP $\leqslant$ 50} &\colhead{FWZP $>$ 50} & \colhead{median FWZP}& \colhead{mean FWZP}\\
\colhead{} & \colhead{} & \colhead{km s$^{-1}$} & \colhead{km s$^{-1}$}  & \colhead{km s$^{-1}$} &\colhead{km s$^{-1}$} & \colhead{km s$^{-1}$} & \colhead{km s$^{-1}$} 
}
\startdata
IRDCs		&	25	 &	1	& 12  	& 11	&1	&20 &23 \\
protostars&	32	 &	5	& 12  	& 14  	&1	&19 &22 \\
\HII		&	45	&	3	& 19  	& 22  	&1	&20 &24 \\
\enddata
%\tablenotetext{1}{The unit of velocity and FWZP is km s$^{-1}$}
\end{deluxetable*}

\begin{deluxetable*}{ccccccc}
\tabletypesize{\scriptsize}
\tablecolumns{7}
\tablewidth{0pc}
\tablecaption{The SiO column densities and abundances.  
\label{tab:NSiO}}
\tablehead{
\colhead{Source groups} & \colhead{$N(\rm SiO)_{min}$} & \colhead{$N(\rm SiO)_{max}$} & \colhead{$N(\rm SiO)_{average}$} &
\colhead{$N(\rm SiO)_{median}$}  & \colhead{$X(\rm SiO)_{average}$}& \colhead{$X(\rm SiO)_{median}$}\\
\colhead{} 				& \colhead{(cm$^{-2}$)} & 
\colhead{(cm$^{-2}$)} & \colhead{(cm$^{-2}$)}  & 
\colhead{(cm$^{-2}$)} & \colhead{}
}
\startdata
IRDCs		&	2.8$\times 10^{11}$	&	3.9$\times 10^{12}$	& $1.0 \times 10^{12}$  	& $9.0 \times 10^{11}$  	& $5.8 \times 10^{-11}$ & $4.1 \times 10^{-11}$\\
protostars&	9.6$\times 10^{10}$	&	4.7$\times 10^{12}$	& $1.2 \times 10^{12}$  	& $6.1 \times 10^{11}$  	& $2.3 \times 10^{-11}$ & $2.0 \times 10^{-11}$\\
\HII		&	$2.4 \times 10^{11}$	&	$2.9 \times 10^{13}$	& $4.4 \times 10^{12}$  	& $2.2 \times 10^{12}$  	& $3.8 \times 10^{-11}$ & $2.7 \times 10^{-11}$\\
All sources	&	9.6$\times 10^{10}$	&	2.9$\times 10^{13}$	& $2.5 \times 10^{12}$  	& $9.3 \times 10^{11}$  	& $4.0 \times 10^{-11}$ & $3.1 \times 10^{-11}$\\
\enddata
%\tablenotetext{}{}
\end{deluxetable*}

%%%%%%%%%%%%%%%%%
%FIGURES
%%%%%%%%%%%%%%%%%

\clearpage

%%%%%%%%%%%%%%%%%
%APPENDIX
%%%%%%%%%%%%%%%%%
\appendix

\section{Appendix information}

%%%%%%%%%%%%%%%%%
%  Figure
%%%%%%%%%%%%%%%%%
\begin{figure} % Example image
\center
\subfloat[]{
\includegraphics[scale=1.]{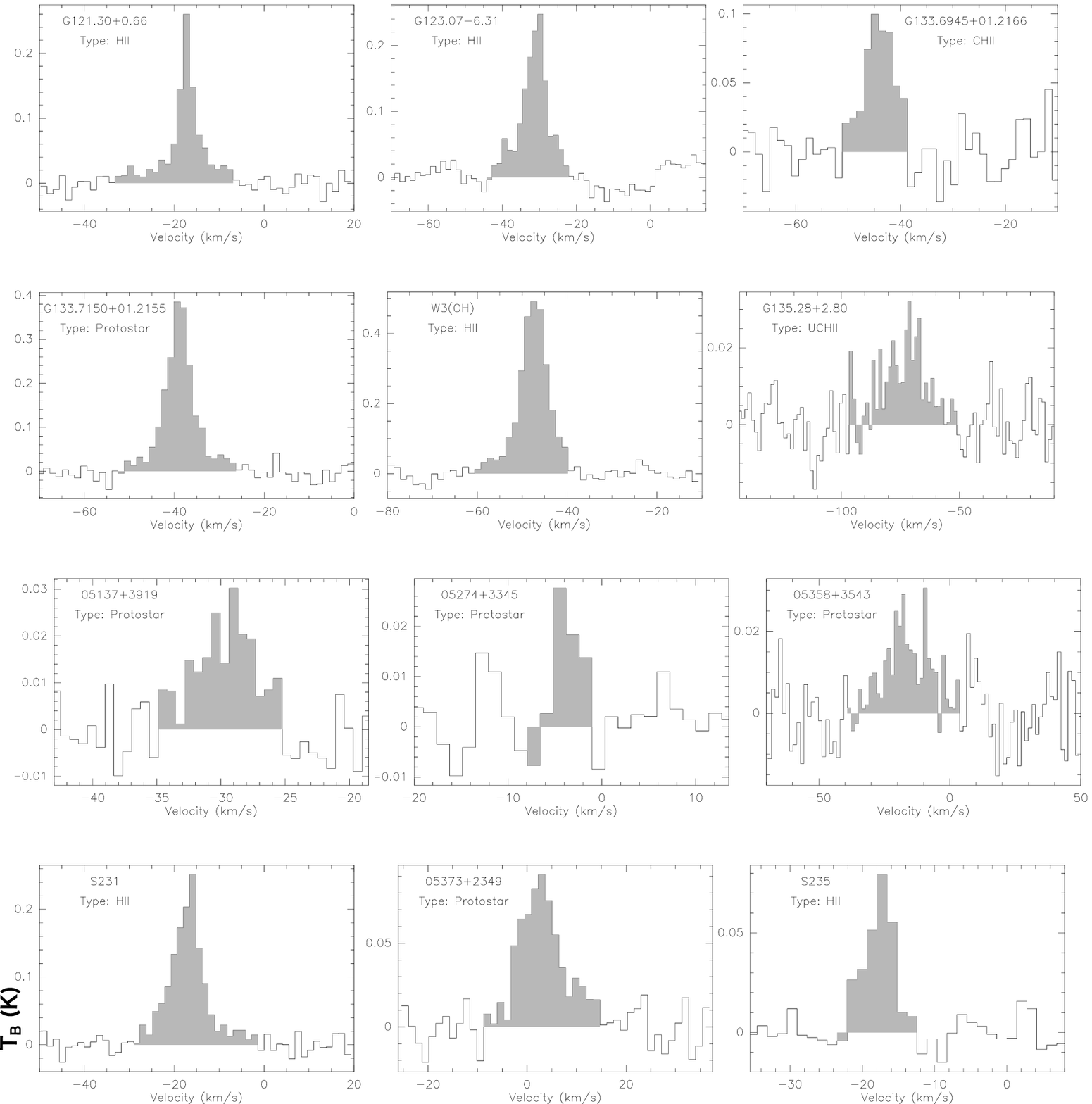}
}\\
\caption{Spectral line of the SiO 5-4 transition in each source. The source name and type are presented in each panel.}
\label{fig:spec}
\end{figure}

\clearpage
\begin{figure}\ContinuedFloat % Example image
\center
\subfloat[]{
 \includegraphics[scale=1.]{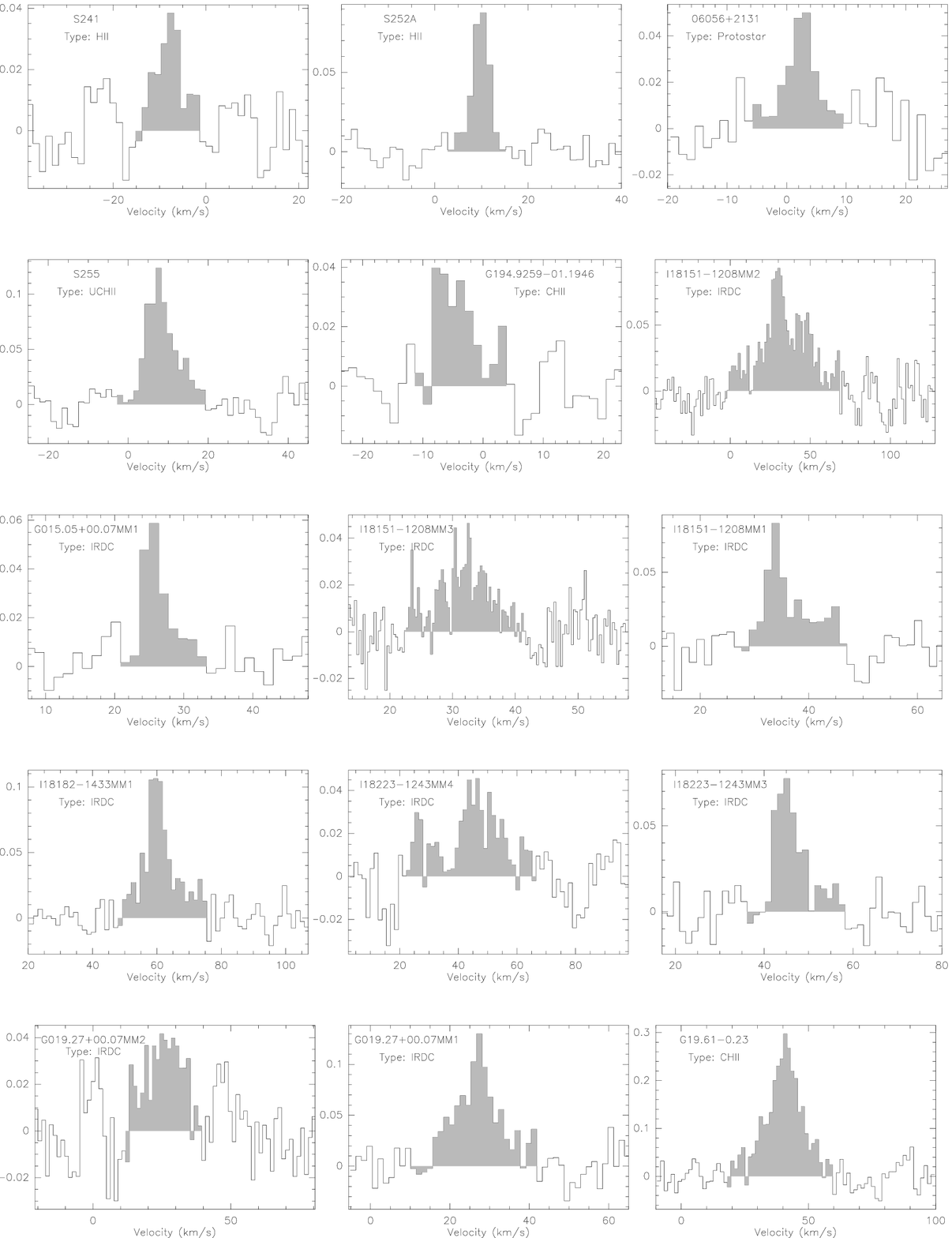}
}\\
\caption{Continuation}
\end{figure}

\clearpage
\begin{figure}\ContinuedFloat % Example image
\center
\subfloat[]{
 \includegraphics[scale=1.]{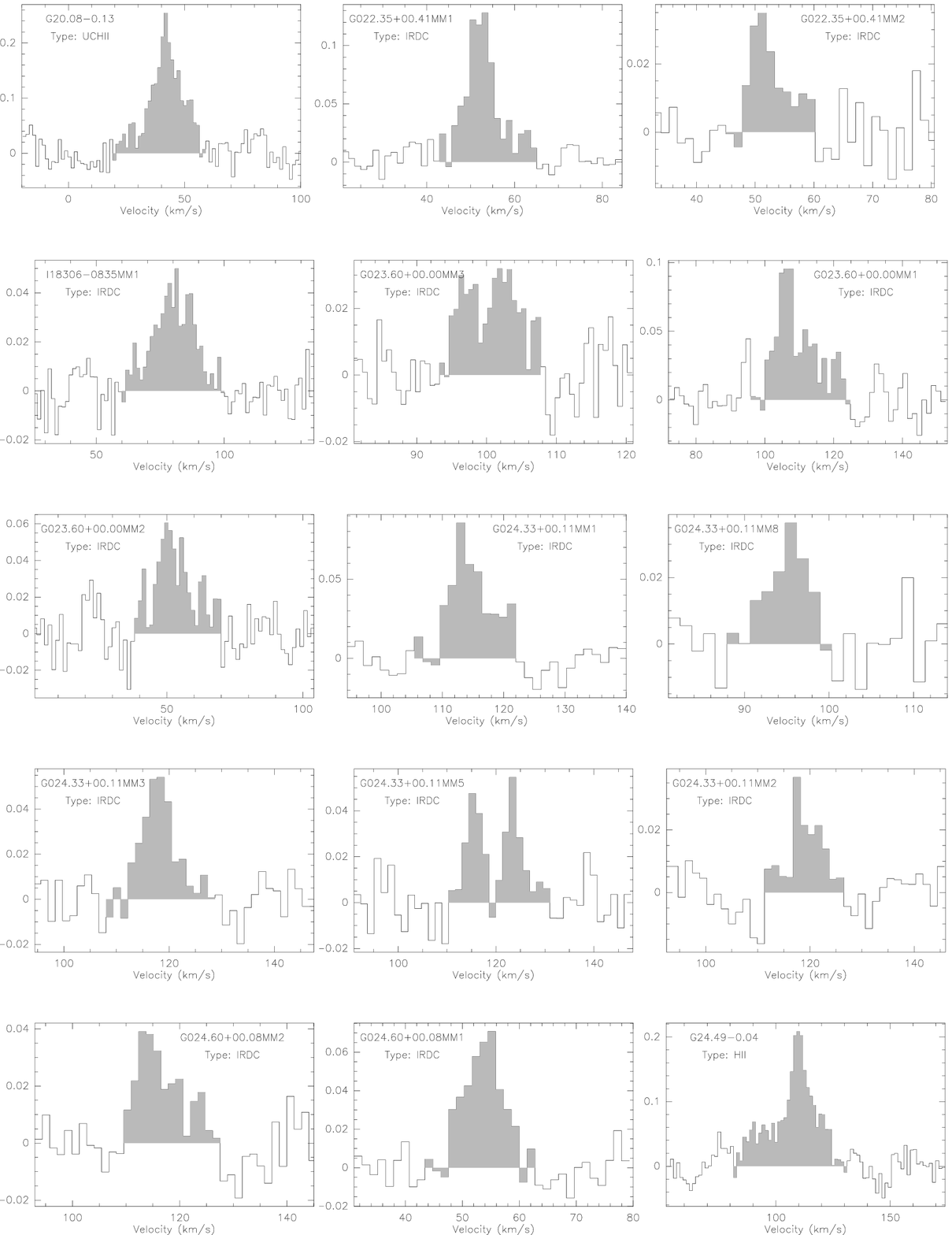}
}\\
\caption{Continuation}
\end{figure}

\clearpage
\begin{figure}\ContinuedFloat % Example image
\center
\subfloat[]{
 \includegraphics[scale=1.]{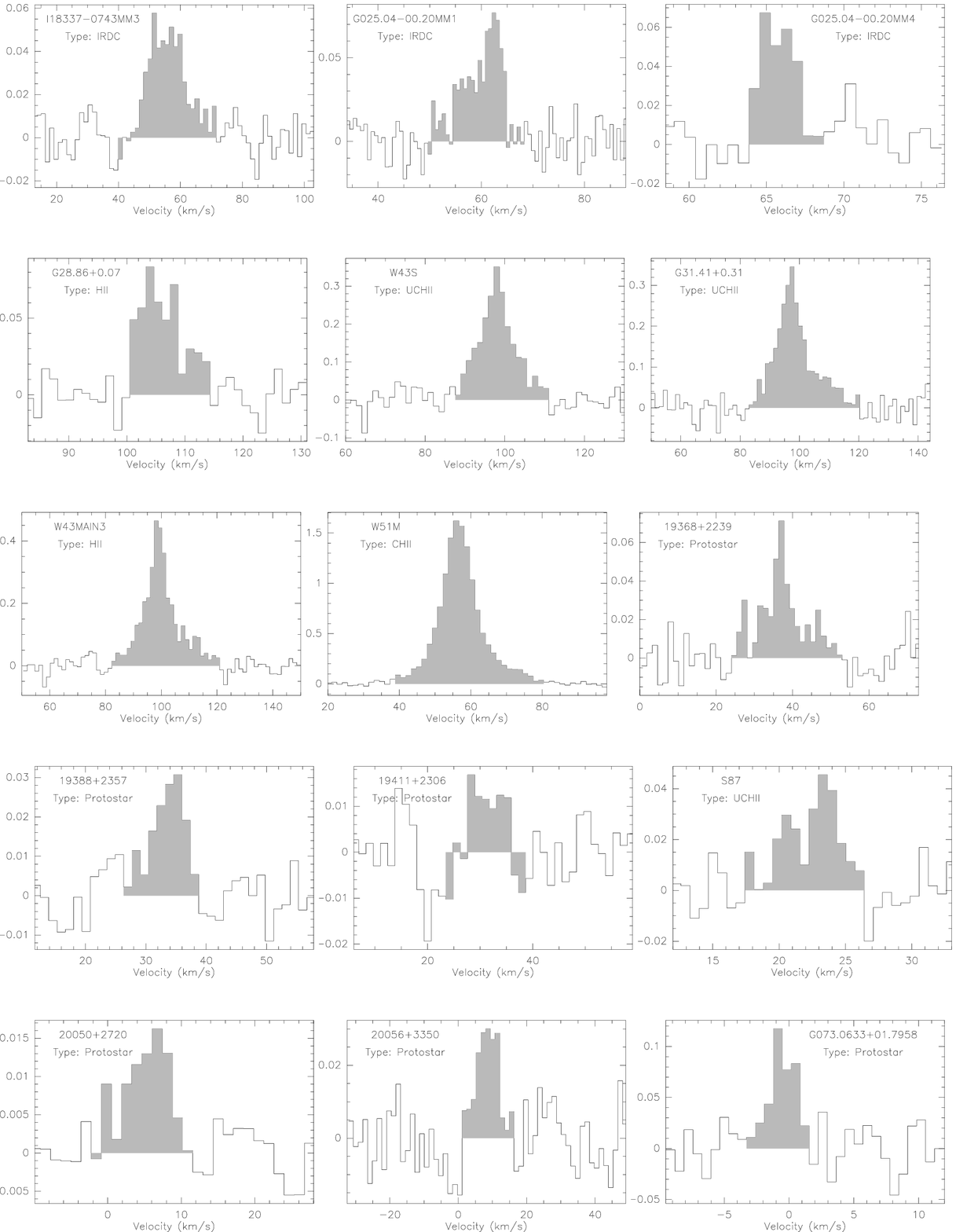}
}\\
\caption{Continuation}

\end{figure}
\clearpage
\begin{figure}\ContinuedFloat % Example image
\center
\subfloat[]{
 \includegraphics[scale=1.]{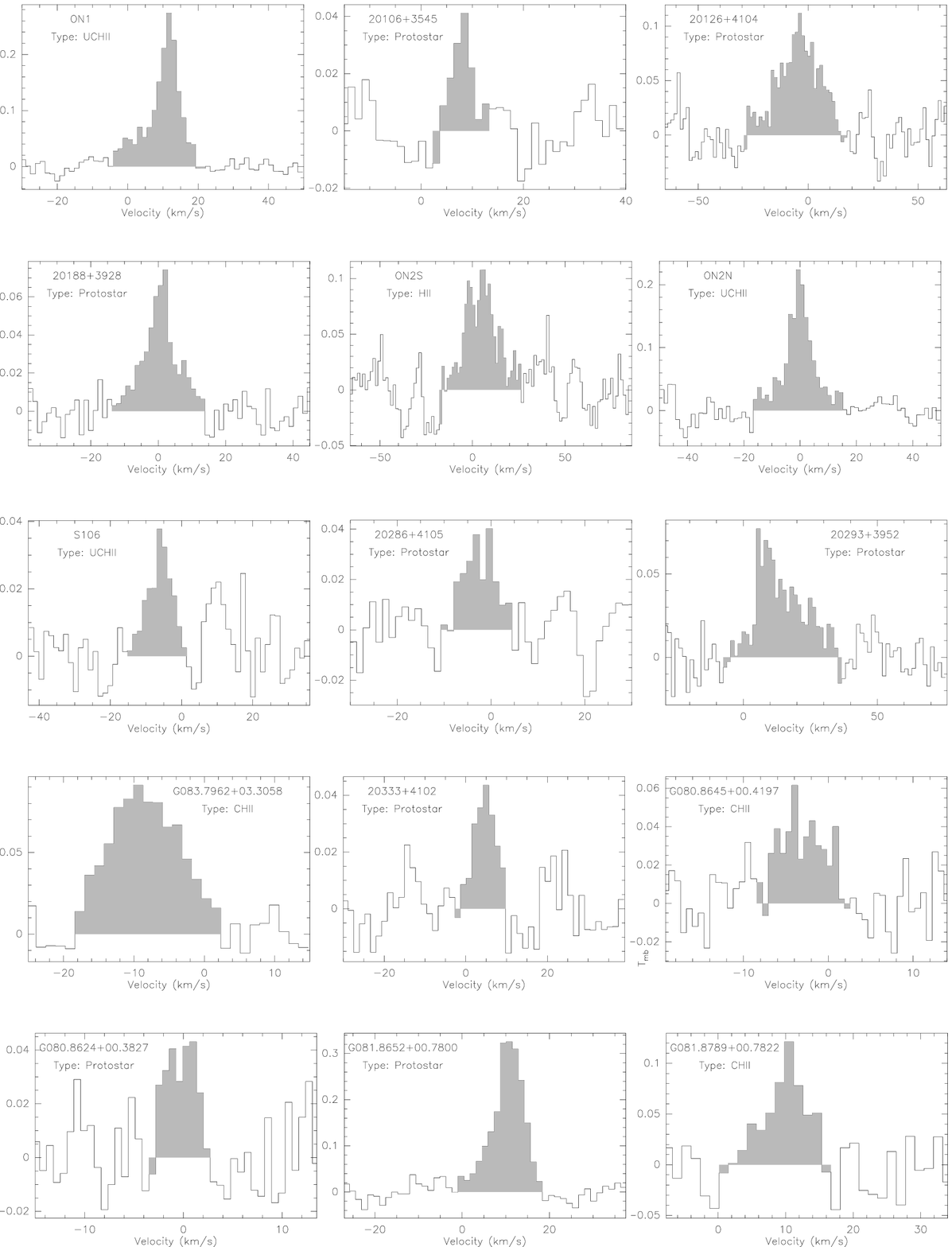}
}\\
\caption{Continuation}
\end{figure}

\clearpage
\begin{figure}\ContinuedFloat % Example image
\center
\subfloat[]{
 \includegraphics[scale=1.]{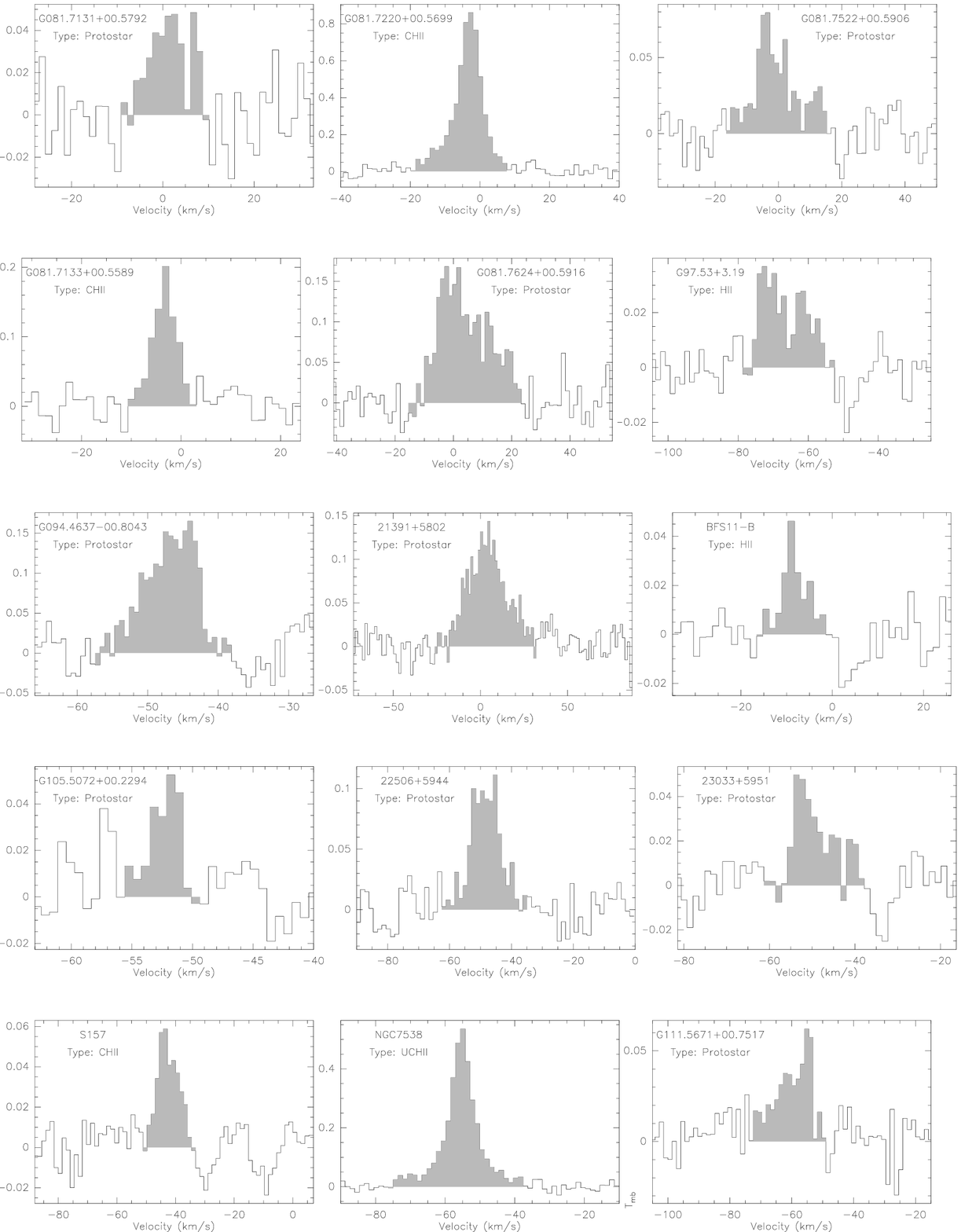}
}\\
\caption{Continuation}
\end{figure}

\clearpage
\begin{figure}\ContinuedFloat % Example image
\center
\subfloat[]{
 \includegraphics[scale=1.]{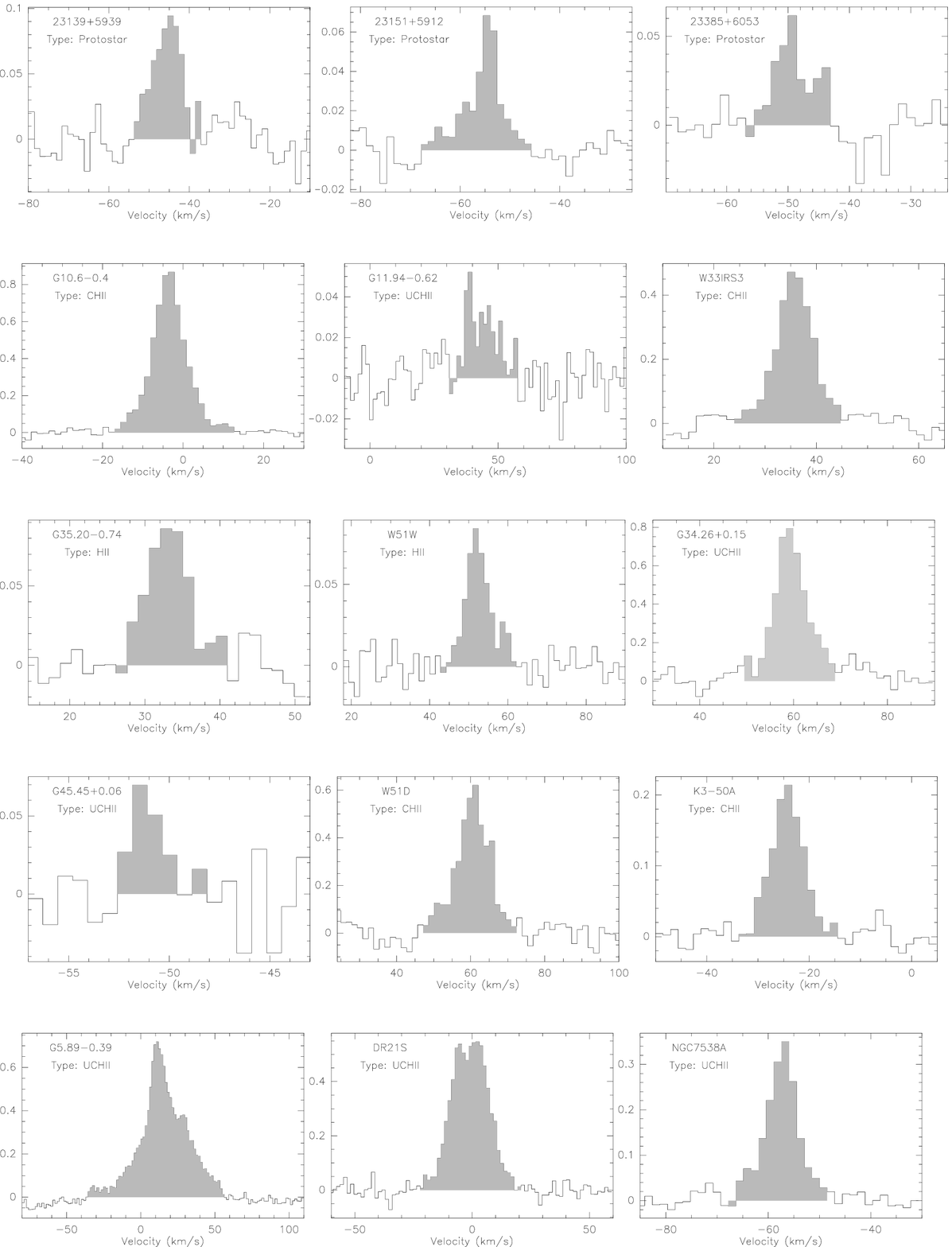}
}\\
\caption{Continuation}
\end{figure}

\label{lastpage}
%\end{CJK*}
\end{document}